\documentclass[aip,amsmath,amssymb,reprint]{revtex4-1}

\usepackage{graphicx}
\usepackage{dcolumn}
\usepackage{bm}

\usepackage[utf8]{inputenc}
\usepackage[T1]{fontenc}
\usepackage{mathptmx}
\usepackage{etoolbox}
\usepackage[dvipsnames]{xcolor}
\usepackage[normalem]{ulem}

\makeatletter
\def\@email#1#2{%
 \endgroup
 \patchcmd{\titleblock@produce}
  {\frontmatter@RRAPformat}
  {\frontmatter@RRAPformat{\produce@RRAP{*#1\href{mailto:#2}{#2}}}\frontmatter@RRAPformat}
  {}{}
}%
\makeatother
\begin{document}

\preprint{AIP/123-QED}

\title[]{Bifurcation Analysis of the Driven FitzHugh-Nagumo Oscillator: Prediction and Experiment}
\author{Edward H. Hellen}
 \email{ehhellen@uncg.edu}
\affiliation{Department of Physics \& Astronomy, University of North Carolina Greensboro, Greensboro, NC 27402 USA
}%


\date{\today}

\begin{abstract}
Bifurcation analysis is applied to the FitzHugh-Nagumo oscillator driven by a sinusoidal source. A numerically generated 2d regime map showing the variety of oscillatory dynamics in the parameter space of source frequency and amplitude agrees well with a map created from analog circuit measurements. Application of the sinusoidal source to the fast variable's first-order differential equation produces an island in the map in which oscillations at the source frequency are unstable and the behavior is dominated by two distinct families of subharmonic limit cycles and by chaos. Previously published maps are portions of the map shown here and are shown to be consistent with it. The more detailed and comprehensive regime map presented here should facilitate the understanding of this foundational system and thereby aid the ongoing research involving more complicated implementations of the Fitzhugh-Nagumo system. 
\end{abstract}

\maketitle

\begin{quotation}
For decades the Fitzhugh-Nagumo equations have been a popular simple model for the action potential in neural and muscle cells. It continues to be used in increasingly sophisticated investigations. One of the early fundamental areas of investigation was the system's response to periodic forcing. Perhaps the most effective way to present the dynamics of a driven excitable oscillator is a regime map in the 2d parameter space of the source's frequency and amplitude. However, the earlier studies did not benefit from today's computing power and bifurcation methods, resulting in publication of partial regime maps. Here a more comprehensive and detailed regime map is presented, revealing an island in which oscillations at the source frequency are unstable. The map is shown to be consistent with earlier published maps, and to agree with a map measured from an analog electronic circuit. The improved map should enhance the understanding of this periodically forced excitable oscillator and inform the design of future experiments using the Fitzhugh-Nagumo model.  

\end{quotation} 

\section{Introduction}
Over the past 60 years the FitzHugh-Nagumo (FHN) equations have achieved wide popularity as a simple model for studying the cell membrane's action potential and the role that excitable-oscillators play in neural signaling and muscle contractility processes.\cite{cebrian2024six} The simplicity of the model is attributable to its use of only two variables--a fast threshold-activated positive feedback variable corresponding to the membrane potential, and a slow inhibitory variable responsible for recovery and a refractory period.\cite{FitzHugh1961,nagumo1962active}

FitzHugh referred to his model as the Bonhoeffer-van der Pol (BVP) model and showed how the BVP model reproduced many of the important features of the more physiologically realistic Hodgkin-Huxley (HH) model\cite{hodgkin1952quantitative}, a 4-variable model used to reproduce action potentials measured in giant squid axon. Fitzhugh's calculations were done using analog computers (analog electronic circuits designed to solve the differential equations). 

At the same time, the group of Nagumo, Arimoto, and Yoshizawa was developing a tunnel-diode based transmission line circuit which demonstrated propagating pulses similar to action potentials in axons. They found that by using FitzHugh's BVP model in a reaction-diffusion model, they could replicate the pulse propagation they measured on their transmission line.\cite{nagumo1962active} The development of the model by FitzHugh and its implementation into a transmission line circuit by Nagumo et.\ al.\ resulted in the model eventually being known as the FHN model. 

It is not an exaggeration to say that there has been an explosion of research using the FHN model. Topics include coupled oscillators, collective behaviors, noise induced effects, multistability, chaos, bursting, stochastic resonance, chimeras, pacing, 2-d wave propagation, memrister variants of FHN, and so on.\cite{rajasekar1988period,feingold1988phase,Alexander1990,SATO1992243,nomura1993bonhoeffer,Moss1994StochRes,BROWN1995359,CHOU1996,Jurgen1997,rajasekar1997control,barnes1997numerical,coombes2000period,kostova2004FitzHugh,tsuji2004design,volkov2005,croisier2007continuation,shimizu2012complex,takahashi2018mixed,zhang2023bifurcation,sakaguchi2023suppression,bosco2024influence,inaba2024nested} An extensive list of references is provided in a recent review.\cite{cebrian2024six}

One of the natural areas of inquiry was how the FHN oscillator responds when driven by an applied periodic source. Dependence on amplitude and frequency of the source are both of interest, as is application of the periodic source to both oscillatory (self-exciting) and excitable (resting at stable fixed point) FHN oscillators. Several studies have addressed these questions.\cite{rajasekar1988period,feingold1988phase,SATO1992243,nomura1993bonhoeffer,BROWN1995359,doi1995global,CHOU1996,barnes1997numerical,coombes2000period,tsuji2004design,croisier2009bifurcation,shimizu2012complex,takahashi2018mixed,inaba2020nested,inaba2024nested} Decades prior to FitzHugh and Nagumo's research, Van der Pol and van der Mark subjected a relaxation oscillator to a sinusoidal source and measured the dependence of the oscillator's pulse period on the source period.\cite{van1927frequency} Instead of varying the source frequency, they did essentially the same thing by varying the oscillator's natural period by adjusting its RC time constant. They found the subharmonic ``period adding" behavior that can occur when the source frequency exceeds a nonlinear oscillator's natural frequency. 

A case can be made that application of a periodic source to a FHN oscillator is a foundational system. Yet it is apparently not easy to find in the literature a comprehensive study of this relatively basic system which includes a detailed 2d regime map showing the dynamical behaviors in the parameter space of source frequency (or period) and amplitude. Portions of maps have been shown,\cite{feingold1988phase,SATO1992243,barnes1997numerical,nomura1993bonhoeffer,doi1995global,coombes2000period,tsuji2004design,croisier2009bifurcation} however a more comprehensive map should be available. Such a map could be informative to the design and understanding of more complex systems, including those involving networks of coupled FHN oscillators and stochastic resonance. The goal of this project is to provide a more detailed regime map covering a larger range in parameter space than is currently available. 

Here we compare a map created from bifurcation analysis methods to the map made from measurements on an analog circuit designed and constructed to solve the system in question. Comparison of the numerically simulated map with actual data is important since it provides confidence in the validity of the numerical bifurcation analysis methods. 

This paper is intended to provide a comprehensive presentation of the dynamics found in the basic system of a sinusoidal source applied to an excitable FHN oscillator characterized by common parameter values. The rich variety of behavior presented in the 2d parameter space of the source frequency and amplitude may help to inform experimentalists in their design and understanding of more complex FHN systems. As such, the paper is not intended to be mathematically rigorous, and describes only the bifurcations needed to account for the observed behaviors. Also, well-established material such as location and stability of fixed points, is described only to the extent relevant to this work. 

\section{FitzHugh-Nagumo Equations}
The FitzHugh-Nagumo equations used here are 
\begin{subequations}
\begin{align} 
\frac{du}{dt}= &(1-u)u(u-a)-v+i+c\:\text{cos}(2\pi ft)\\
\frac{dv}{dt}= & \epsilon (u-bv)
\end{align}
\label{nonauton-eqns}
\end{subequations}
where the sinusoidal source has amplitude $c$ and frequency $f$. Equations \eqref{nonauton-eqns} are non-autonomous due to the source's time dependence. The amplitude and frequency of the source term are the parameters of interest to vary in this study. The other parameter values used here are $a=0.25$, $b=1$, $\epsilon=0.01$, and $i=0.1$. As shown below, this value of $i$ results in a stable fixed point when there is no sinusoidal source. 

Fitzhugh's original equations\cite{FitzHugh1961} use a cubic polynomial of the fast variable with zeros at $u=0,\text{ and }\pm\sqrt{3}$. The cubic term used in Eq.\ \eqref{nonauton-eqns} is sometimes preferred mathematically because its zeros at $u=0$, $a$, and 1 produce an excitation pulse with a resting value near zero and a pulse maximum near 1 when using common parameter values. The threshold activated positive feedback occurs when $a<u<1$. This positive feedback competes with the negative feedback produced by the $-v$ term from the slow inhibitory variable to determine the overall feedback.  

The FHN system has already been thoroughly analyzed.\cite{cebrian2024six} Here we briefly cover topics relevant to this work. To start, we look at the dependence of the FHN oscillator dynamics on the constant source term $i$ so that we can choose $i$ such that the oscillator is at a stable fixed point close to zero when there is no sinusoidal source. The $i$-value which corresponds to a fixed point $u^*$ of Eq.\ \eqref{nonauton-eqns} (with $c=0$) is readily found to be
\begin{equation}
    i^*=\frac{u^*}{b}-(1-u^*)u^*(u^*-a).
\label{i-fixed}
\end{equation}
Stability of a fixed point is determined using standard methods.\cite{CHOU1996} A fixed point is stable if the real parts of the two eigenvalues of the Jacobian of Eq.\ \eqref{nonauton-eqns} (with $c=0$) are negative, and unstable if one or both of the real parts are positive. Calculating the Jacobian, its eigenvalues, and then setting the real parts of the eigenvalues to zero gives 
\begin{equation}
    -3u^2+2(1+a)u-a=\epsilon b.
\end{equation}
The two solutions of this equation are  
\begin{equation}
    u^{\pm}=\frac{1}{3}\left(1+a\pm\sqrt{1-a+a^2-3\epsilon b}\right).
\end{equation}
These two values are used in Eq.\ \eqref{i-fixed} to find the two $i$-values where the fixed point changes stability, at $i^-=0.1355$ and $i^+=0.6168$. The stability analysis finds that the fixed point is stable for $i<i^-$ and $i>i^+$. 

\begin{figure} [h]
\includegraphics[width=3.2 in]{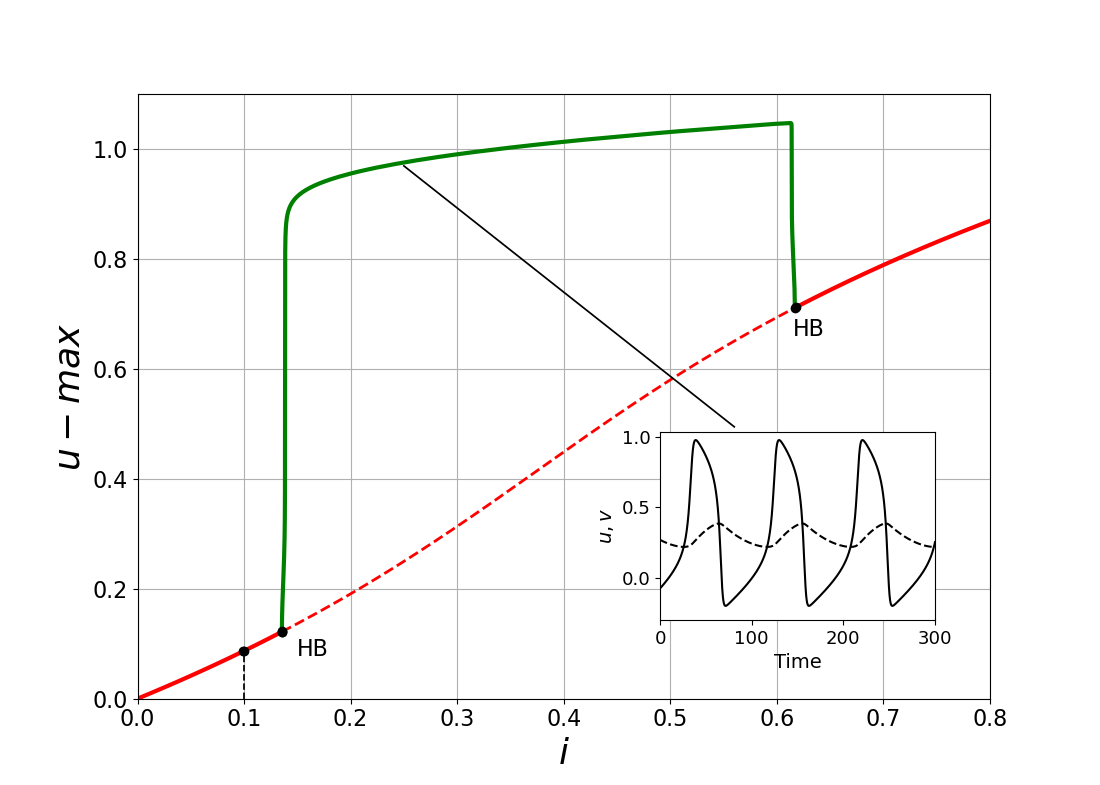}
\caption{Numerical $i$-continuation bifurcation plot showing the maximum value of $u$ for no sinusoidal source. Red is fixed point, green is limit cycle. Solid (dashed) line indicates stable (unstable). Inset shows stable limit cycle time-series of $u$ (solid) and $v$ (dashed) at indicated value $i=0.25$. Value $i=0.1$ is used for this paper. $\epsilon=0.01$, $a=0.25$, and $b=1$.}
\label{i-contin}  
\end{figure}

Figure \ref{i-contin} is an $i$-continuation bifurcation plot showing how the FHN oscillator dynamics depend on parameter $i$.
The red line shows the fixed point $u^*$ which becomes unstable (dashed) between the Hopf bifurcations (HB) located at the $i^{\pm}$ values calculated above. The green line shows the maximum $u$-value of the stable limit cycle (LC). The inset shows the time-series of the limit cycle that occurs when $i=0.25$. In this work we are interested in oscillations caused when the sinusoidal source term is applied to an otherwise resting oscillator. Choosing $i=0.1$ produces a stable fixed point $u^*=0.087$ as indicated in Fig.\ \ref{i-contin}. 

In order to choose an interesting frequency range for investigation, it is helpful to first deduce a natural oscillation period for Eq.\ \eqref{nonauton-eqns} in the absence of the sinusoidal source. One approach is to consider the periods of the LC which exists between the HB points in Fig.\ \ref{i-contin}. Numerical simulation finds that in the range $0.15<i<0.6$ the dimensionless oscillation period is from about 85 to 115. This result is consistent with the second approach which is to consider the characteristic recovery time $1/\epsilon=100$ for the slow variable $v$. Thus, a period of 100 suggests that for sinusoidal forcing, frequencies $f<<0.01$ will produce quasi-steady-state responses and frequencies $f>>0.01$ will be too fast for the system to respond. These considerations suggest that dimensionless frequencies in the vicinity of 0.005 to 0.05 are of interest in this study. 

Electronic circuit equations are obtained from Eq.\ \eqref{nonauton-eqns} by converting from dimensionless quantities to voltages in volts and time in seconds. Variables $u$ and $v$ and parameters $a$ and $i$ are converted to voltages by scaling each voltage by 10 volts,   
\begin{equation}
u=\frac{V_u}{10} \quad v=\frac{V_v}{10} \quad a=\frac{V_a}{10} \quad i=\frac{V_i}{10}.
\label{dimen}
\end{equation}
Time is scaled by an $RC$ time constant, $t\rightarrow t/(RC)$. Using these conversions in Eq.\eqref{nonauton-eqns} results in
\begin{subequations}
\begin{align} 
RC\frac{dV_u}{dt}= &\left(\frac{10-V_u}{10}\right) V_u\left(\frac{V_u-V_a}{10}\right)-V_v+V_i+V(t)\\
RC\frac{dV_v}{dt}= & \epsilon (V_u-bV_v)
\end{align}
\label{circuit eqns}
\end{subequations}  
where $t$ is dimensionless in Eq.\ \eqref{nonauton-eqns} and has units of seconds in Eq.\ \eqref{circuit eqns}. $V(t)$ is the sinusoidal voltage source. We note that the parameters $\epsilon$ and $b$ are coefficients and do not undergo conversion.  

Some of the bifurcation analysis methods are more easily suited to autonomous systems. Therefore, Eq.\ \eqref{nonauton-eqns} is converted to an autonomous form by including two additional variables $x(t)$ and $y(t)$. The differential equations for these variables are constructed so that $x$ and $y$ are sinusoidal with frequency $f$ and unity amplitude.\cite{ermentrout,auto-07p} The autonomous version of system \eqref{nonauton-eqns} is then
\begin{subequations}
\begin{align} 
\frac{du}{dt}= &\:(1-u)u(u-a)-v+i +c x\\
\frac{dv}{dt}= &\: \epsilon(u-bv) \\
\frac{dx}{dt}= & x(1-x^2-y^2)-2\pi f y\\
\frac{dy}{dt}= &\: y(1-x^2-y^2)+2\pi f x.
\end{align}
\label{auton-eqns}
\end{subequations}

The commonly used periodic sources are either single polarity periodic pulses inspired by cell membrane action potentials, or sinusoids routinely used in engineering and physics for studies of system response. Here we use sinusoidal. However, we are able to make useful comparisons of our results to previous studies which used periodic pulses. 

\section{Bifurcation Analysis Method}
Bifurcation analysis of Eq.\ \ref{auton-eqns} was done using XPPAUT \cite{ermentrout}. 
This software tool is used to find the limit cycles (LC), their bifurcation points, Poincar\'e plots, and tracings of the bifurcation points in 2d parameter space thereby producing regime maps showing the types of LCs and oscillatory dynamics. Time-series integration was done using an adaptive step-size 4\textsuperscript{th}-order Runge-Kutta routine. Lyapunov exponents were calculated using the method of Wolf et.\ al.\ \cite{wolf1985} The bifurcation analysis produces bifurcation continuation plots and the 2d regime maps. The bifurcations for Eq.\ \ref{auton-eqns} are period-doublings (PD) and limit points (LP). Here we describe the bifurcation points and their relation to the bifurcation continuation plots. 

PD points occur on a bifurcation continuation plot at parameter values where a stable LC becomes unstable and a LC with double the period is created. The main LC of Eq.\ \ref{auton-eqns} is period-1 (P1), meaning it has the same period as the sinusoidal source. This LC exists as either stable or unstable over the entire parameter space. Figure 
\begin{figure} [h]
\includegraphics[width=3.2 in]{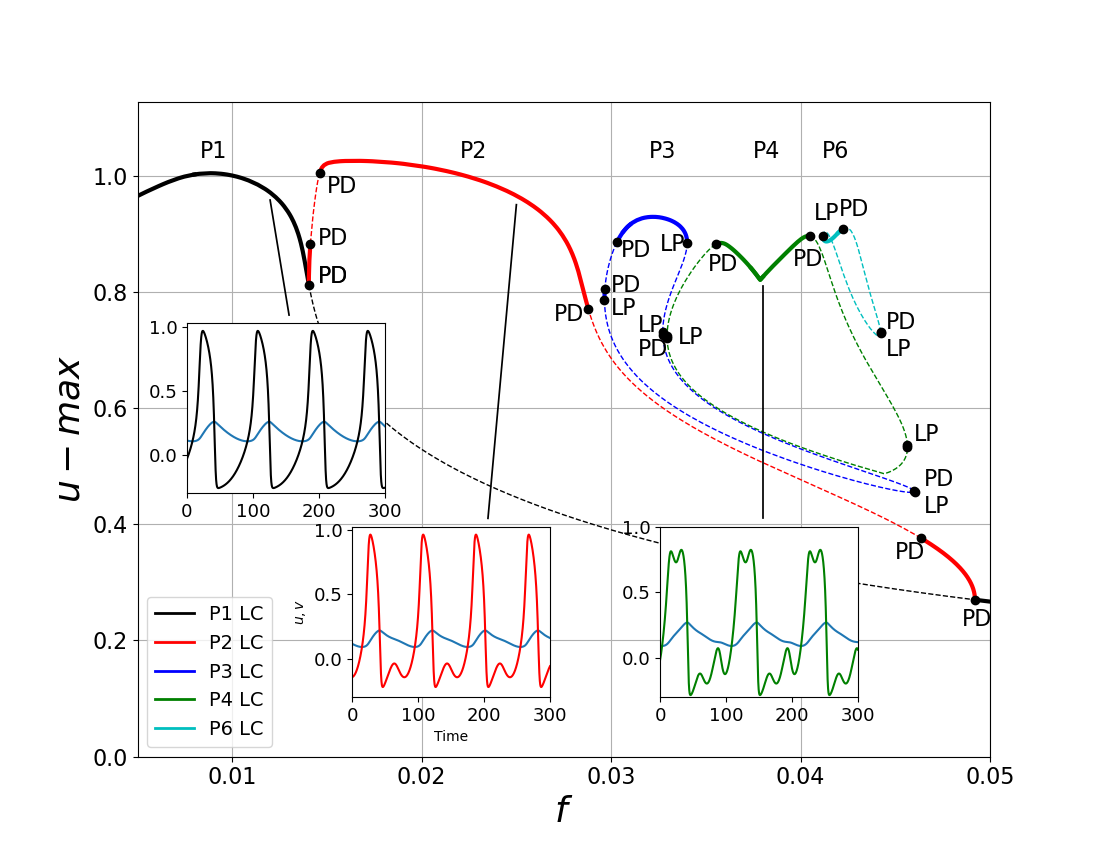}
\caption{Numerical $f$-continuation bifurcation plot showing the maximum value of $u$ for the P1, P2, P3, P4, and P6 LCs. $\epsilon=0.01$, $a=0.25$, $b=1$, $i=0.1$, and $c=0.05$. Solid (dashed) line indicates stable (unstable) LC. PD and LP bifurcation points are indicated. Insets show stable LC time-series of $u$ and $v$ (blue) for P1 at $f=0.012$, P2 at $f=0.025$ and P4 at $f=0.038$.}
\label{f-contin}  
\end{figure}
\ref{f-contin} shows the $f$-continuation bifurcation plot for P1 LC and the higher period LCs, P2, P3, P4, and P6 for source amplitude $c=0.05$. Many PD points are seen where the LCs change stability. Solid (dashed) line indicates stable (unstable) LC. The plot shows how $u$-max (the maximum value of the LC's variable $u$) depends on parameter $f$. The insets in Fig.\ \ref{f-contin} show time-series of $u$ and $v$ for the P1, P2, and P4 LCs at the indicated $f$-values. 

LPs (also called saddle-node bifurcations) occur on a bifurcation continuation plot at parameter values where a stable LC branch and an unstable LC branch extending together in parameter space reach an extreme parameter value (the LP) where they collide and disappear. Another description is that a LC branch reverses its direction along the parameter-axis at the LP, with concomitant change in stability. 

Figure \ref{f-contin} shows that the endpoints of the stable LCs are either PD or LP bifurcations. P1 is stable at low frequencies up to the PD at 0.0140, then returns to stable at the PD at 0.0492. P2 is stable between the PD points at 0.0146 and 0.0288. P3 is stable between the PD at 0.0303 and the LP at 0.0340. P4 is stable between PDs at 0.0355 and 0.405. P6 is stable between the LP at 0.0412 and the PD at 0.0422. For other values of source amplitude $c$, the borders of the stable LCs may switch from PD to LP and vice versa. These switches are found by the tracings described next. 

The LP and PD bifurcation points detected in the parameter continuation plots are the starting points for tracing their paths in the 2d $f$-$c$ parameter space. The paths are the borders for different dynamical regimes. Creation of these dynamical regime maps is a goal of this study because of their usefulness in presenting how the system Eq.\ \ref{nonauton-eqns} behaves throughout the $f$-$c$ parameter space.  

The Lyapunov exponents (LE) are useful for detecting regions with no stable LCs, where the oscillatory states are either quasi-periodic or chaotic. Borders for chaotic regions are not detected by the bifurcations tools but are easily identified in LE plots of autonomous systems as parameter values where the maximal LE changes from zero (indicating stable LC) to positive (indicating chaos).     

\section{Circuit}\label{circuit section}
\begin{figure} [h]
\includegraphics[width=3.4 in]{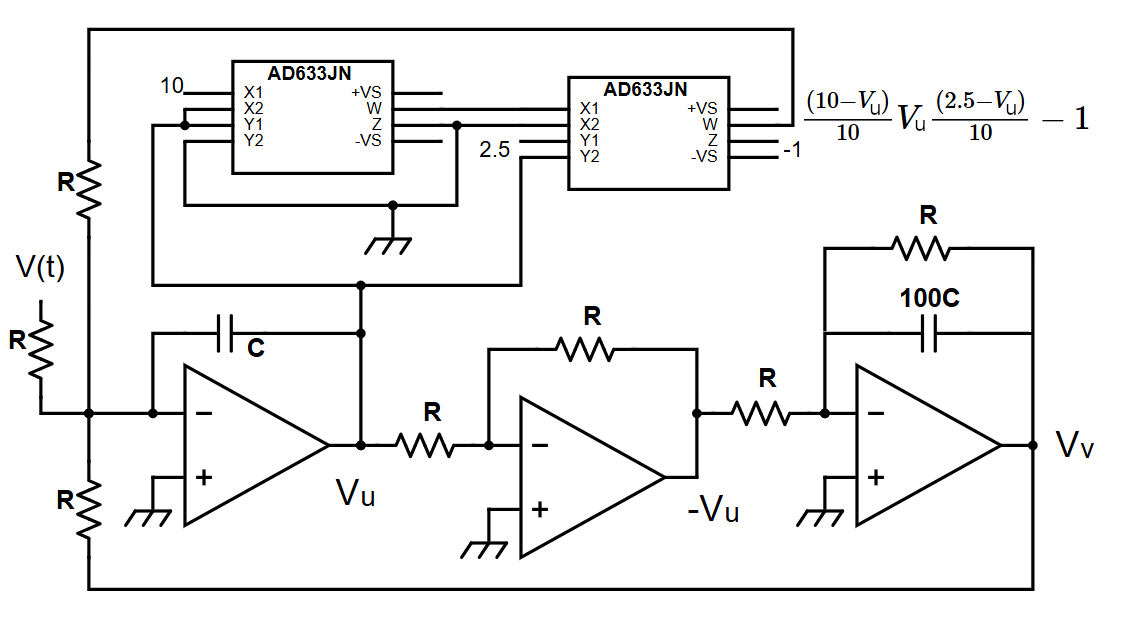}
\caption{Analog circuit for the FitzHugh Nagumo system with sinusoidal source $V(t)$. $R=10^4 \Omega$, $C=10^{-9}$ f.}
\label{circuit}  
\end{figure}

Figure \ref{circuit} shows the op-amp based circuit that was designed and constructed as an analog computer of the FHN system Eq.\ \eqref{nonauton-eqns}. The circuit equations are Eq.\ \eqref{circuit eqns}. Summing the currents into the capacitors of the 1\textsuperscript{st} and 3\textsuperscript{rd} op-amps in Fig.\ \ref{circuit} produces the two differential equations in Eq.\ \eqref{circuit eqns}. The time scale is set by $RC=10^4\Omega\times 10^{-9}\text{f}=10^{-5}$ s.

The multiplication of voltages required in Eq.\ \eqref{circuit eqns} is accomplished using two AD633 multiplier chips (Analog Devices). This chip scales its output voltage by division by 10. The two factors of 10 in the denominator of the cubic term in Eq.\eqref{circuit eqns} account for the scalings from the two voltage multiplications. 

The factor of 100 for $C$ in the negative feedback of the 3\textsuperscript{rd} op-amp accounts for $\epsilon=0.01$. The voltages $-1$, $2.5$, and 10 volts are applied to the indicated inputs of the AD633 chips to produce the cubic term shown at the output of the second AD633. Op-amps are OPA2228 (Texas Instruments). The op-amps and the AD633 chips are powered by $\pm 15$ volts.

\section{Results}
\subsection{Numerical Bifurcation Analysis}
\begin{figure} [h]
\includegraphics[width=3.4 in]{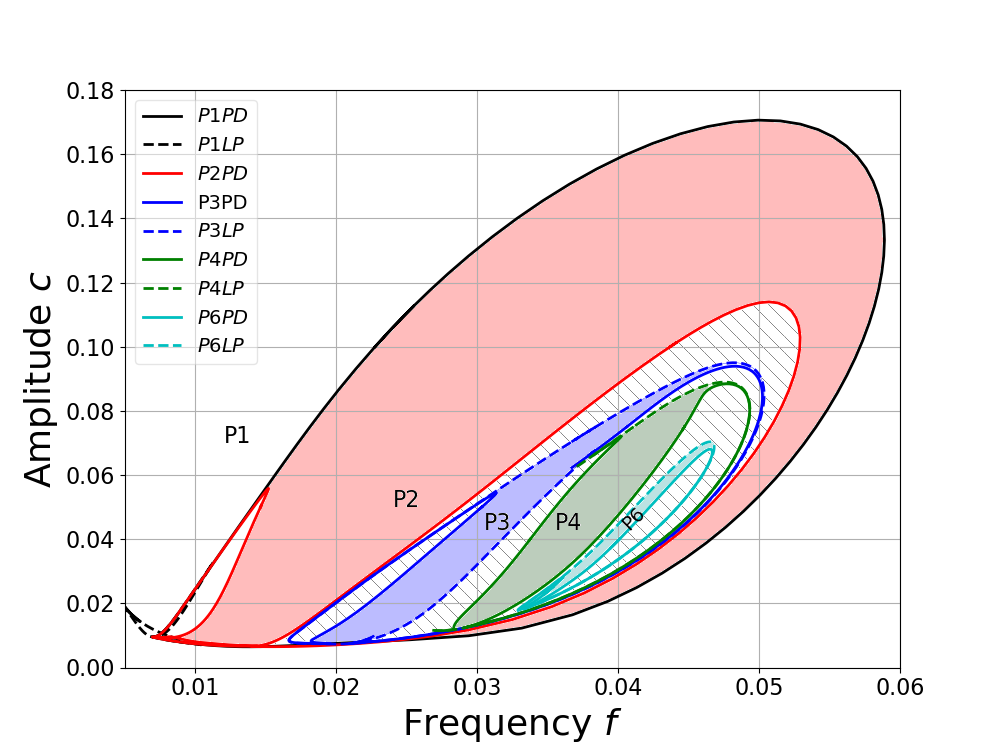}
\caption{Numerical $f$-$c$ regime map for $\epsilon=0.01$, $a=0.25$, $b=1$, and $i=0.1$. P1-P4 and P6 LC regions are indicated. Frequency $f$ and source amplitude $c$ are dimensionless. LP lines are dashed, PD lines are solid. Chaos indicated by diagonal lines.}
\label{fc-regime map}  
\end{figure}

Figure \ref{fc-regime map} is the numerical dynamical regime map showing locations of the limit cycles and complex oscillations of the sinusoidally driven FHN system in the 2-dim $f$-$c$ parameter space. The map consists of PD and LP lines. For clarity, only the PD and LP lines which form borders of stable LCs are shown. 
Traversing the map at $c=0.05$ allows comparison with the bifurcations and stable LCs shown in Fig.\ \ref{f-contin}. The P1 to P2 transition occurs from $f=0.0140$ to $0.0146$, P2 to P3 occurs from $f=0.0288$ to $0.0303$, P3 to P4 occurs from $f=0.0340$ to $0.0355$, and P4 to P6 occurs from $f=0.0405$ to $0.0412$. Then over the $f$-span from 0.042 to 0.046, the system returns to P2 and then P1, and the amplitude drops from $\approx0.9$ for P6 LC down to $<0.4$ for P2 and P1 according to Fig.\ \ref{f-contin}. 

The transition regions can be identified as the regions separating the labeled (P1, P2, etc) stable LCs in Figs.\ \ref{f-contin} and \ref{fc-regime map}. Chaos, indicated by the diagonal lines in Fig.\ \ref{fc-regime map}, is shown below to be the dominant behavior in these regions. Small periodic windows occur within the chaos. Interestingly, stable P5 LC does not occur over broad ranges like those shown for the other LCs in Fig.\ \ref{fc-regime map}. 

The region of unstable P1 LC forms an island in the 2-dim $f$-$c$ parameter space. Much of the island is occupied by stable higher period P$n$ LCs. This observation suggests that continuation bifurcations at other $c$-values will be similar to the continuation plot Fig.\ \ref{f-contin} in which the majority of the span across the island is occupied by stable LCs comprised of the higher period LCs.

\begin{figure} [h]
\includegraphics[width=2.5 in]{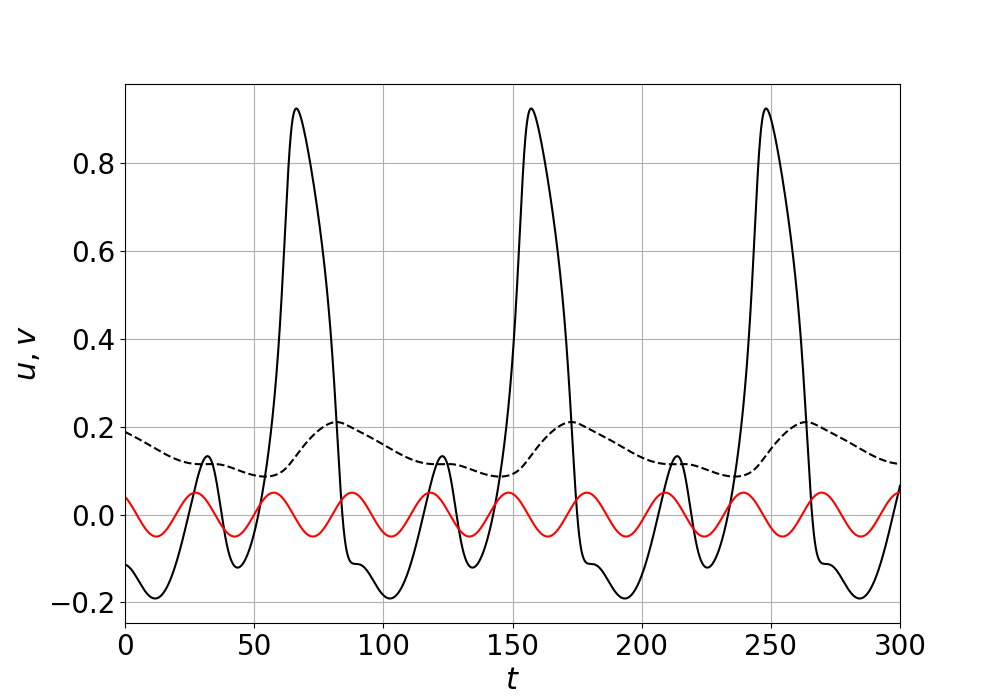}
\caption{Numerical time series of the P3 LC for $f=0.033, c=0.05$. $u$ solid line, $v$ dashed line. Sinusoidal source is shown in red.}
\label{time series P3}  
\end{figure}

Figure \ref{time series P3} shows the time series for P3 LC which occurs at $(f=0.033,c=0.05)$ on the Fig.\ \ref{fc-regime map} map. 
The sinusoidal source is shown in red, which confirms the designation as P3 since three source periods equal one P3 LC period. The structure of the P3 LC is consistent with the P2 and P4 LCs in the insets of Fig.\ \ref{f-contin}, single large peaks separated by smaller oscillations in between and periods close to 100. 

\begin{figure}[htbp]
\includegraphics[width=3 in]{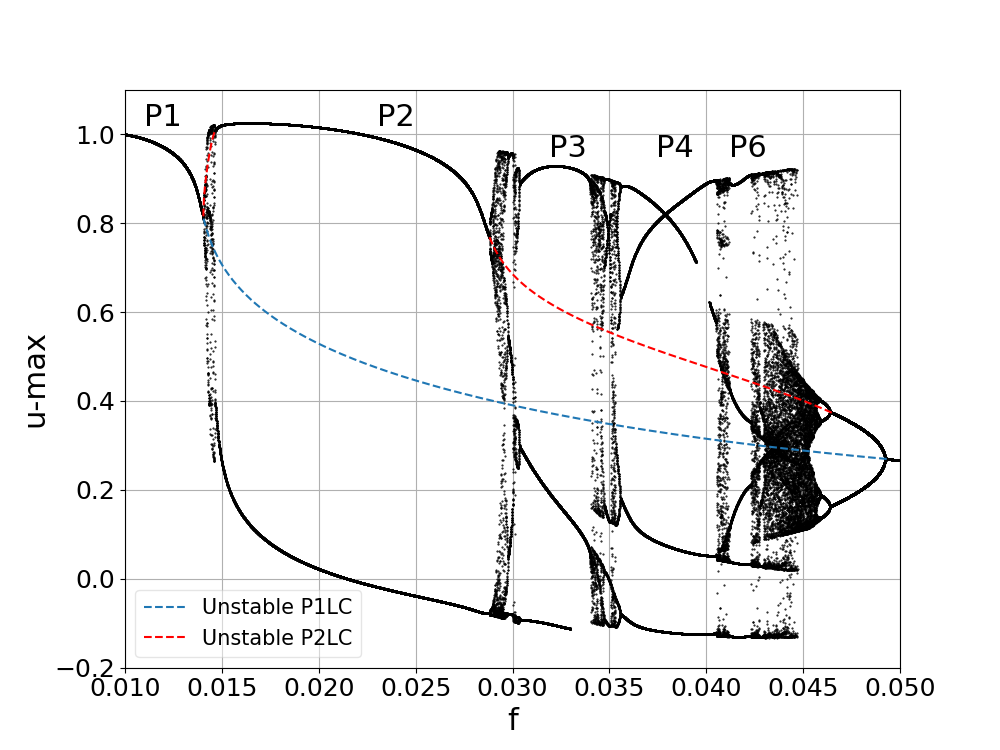}
\caption{Poincar\'e bifurcation diagram for increasing source frequency $f$ with $c=0.05$. Unstable branches of P1 and P2 LCs shown as dashed lines.}
\label{poincare diagram}  
\end{figure}

Figure \ref{poincare diagram} shows a local-maximum based Poincar\'e bifurcation diagram corresponding to the same conditions as the continuation in Fig.\ \ref{f-contin}. The regions of stable LCs separated by regions of complex oscillations in these two figures are in agreement.  
The unstable portions of the P1 and P2 LC branches in the continuations in Fig.\ \ref{f-contin} are shown as dashed lines in Fig.\ \ref{poincare diagram}. The period-doubling cascade which returns the system to stable small amplitude P1 LC at high $f$ is clearly seen. Confirmation of chaos by calculation of Lyapunov exponents is in the next section. 

\begin{figure}[htbp]
\includegraphics[width=3 in]{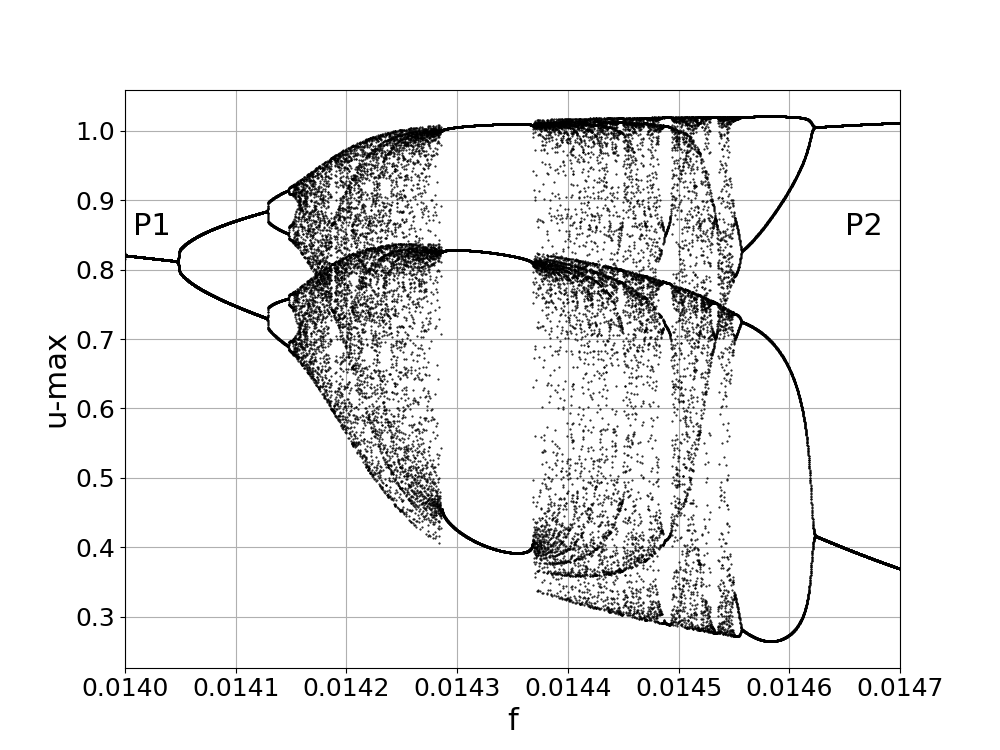}
\caption{Zoom of the Poincar\'e bifurcation diagram showing the complex behavior during the transition from P1 to P2 LC. The P1 LC undergoes a period doubling cascade to chaos which ends in a reverse doubling to P2 LC. A window of period-3 LC becomes stable within the chaos. Source amplitude $c=0.05$.}
\label{poincare zoom diagram}  
\end{figure}

Figure \ref{poincare zoom diagram} shows a zoom of the complex behavior in the transition region between stable P1 and stable P2 LC just below $f=0.015$ in Fig.\ \ref{poincare diagram}. The P1 LC undergoes a period-doubling cascade to chaos, eventually undergoing a reverse-doubling to stable P2 LC. A periodic window of period-3 LC appears inside the chaos.  
This period-3 LC is distinctly different from the P3 LC occupying the much larger area of the map Fig.\ \ref{fc-regime map}. It is part of a family of subharmonic LCs discussed below which have a very different structure than the P2-P6 LCs occupying the large portion of the unstable P1 island.  

\begin{figure} [htbp]
\includegraphics[width=3 in]{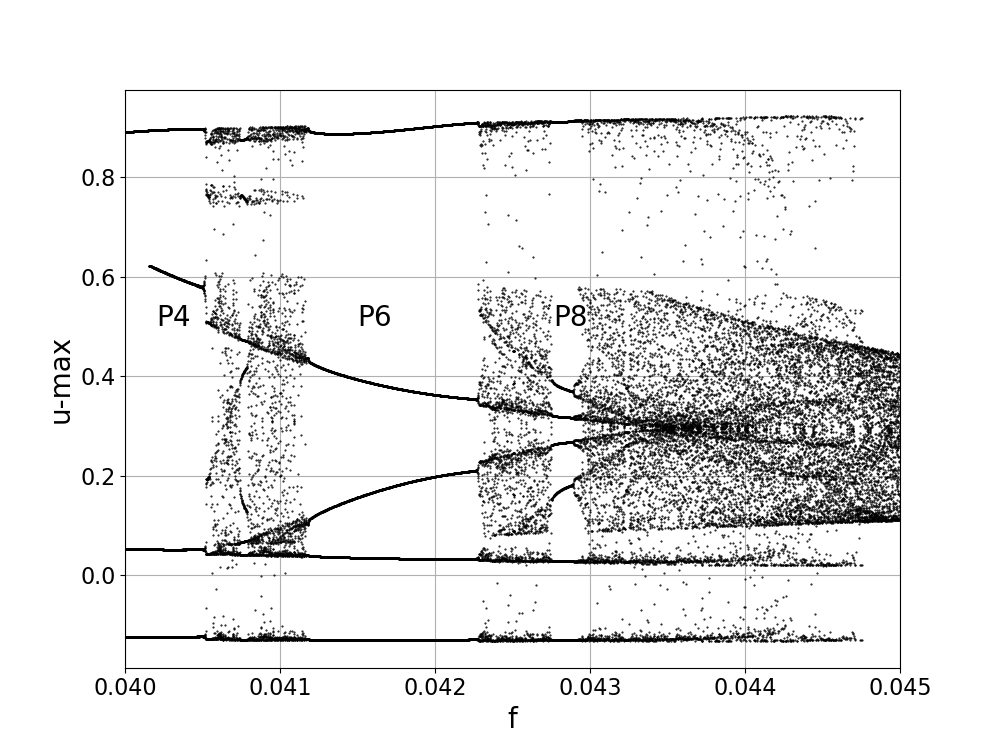}
\caption{Zoom of the Poincar\'e bifurcation diagram showing the P6 and P8 LC regions. $c=0.05$.}
\label{poincare zoom P8}  
\end{figure}

\begin{figure} [h]
\includegraphics[width=1.6 in]{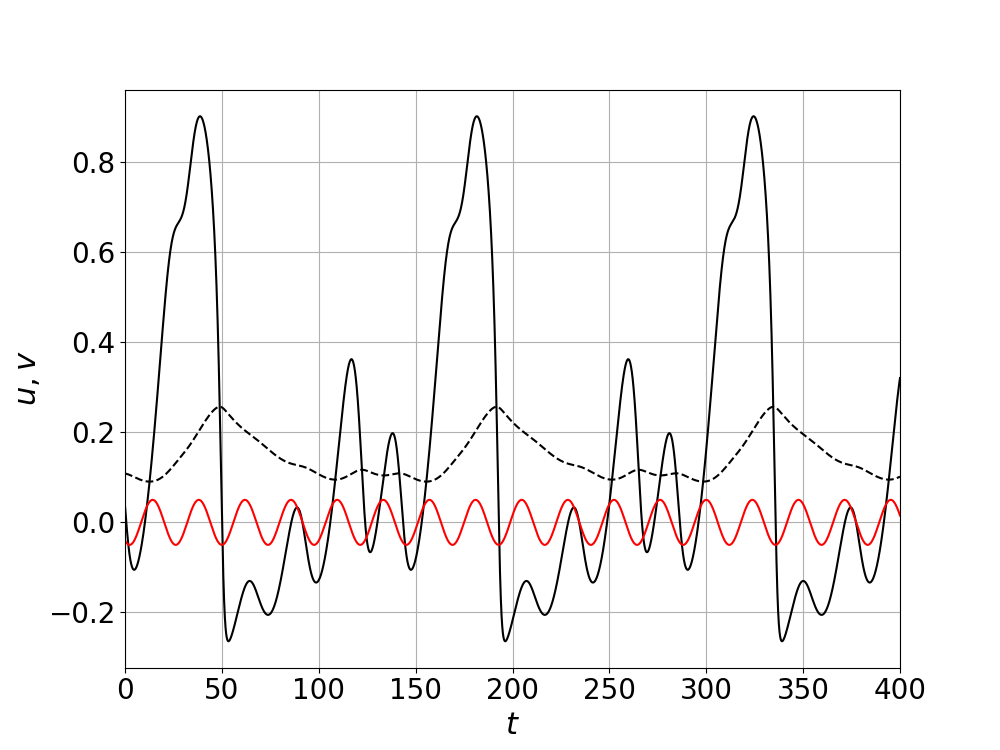}\quad
\includegraphics[width=1.6 in]{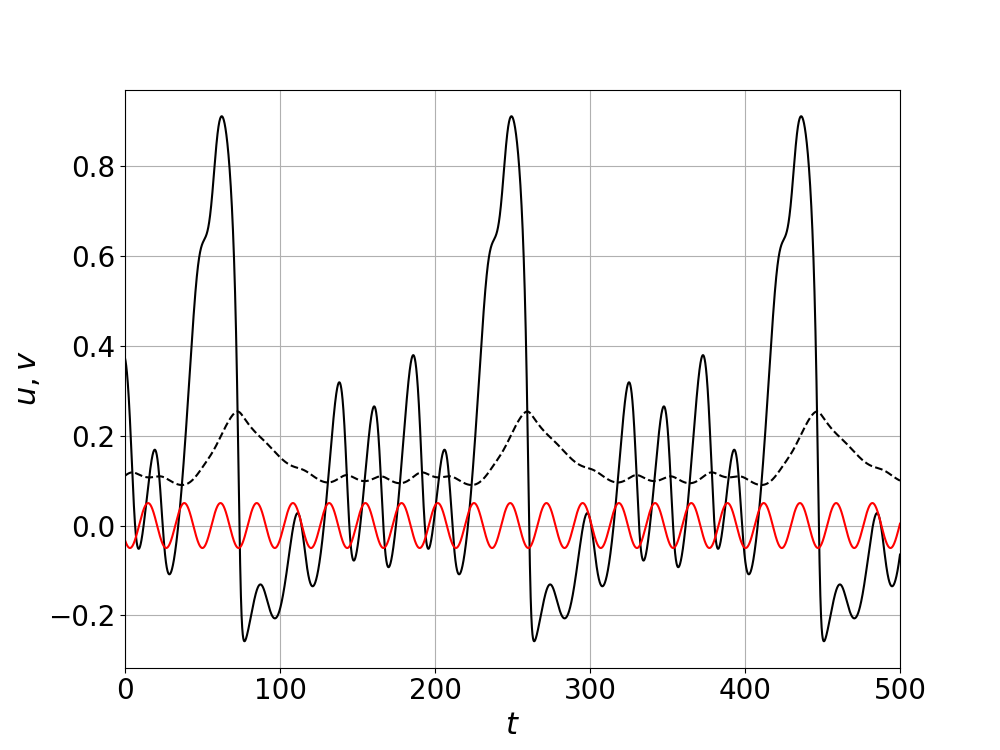}
\caption{Numerical time series of the P6 ($f=0.042$) and P8 ($f=0.0428$) LCs for $c=0.05$. $u$ solid line, $v$ dashed line. Sinusoidal source is shown in red.}
\label{time series P6 P8}  
\end{figure}

The Poincar\'e plot in Fig.\ \ref{poincare zoom P8} is a zoom of the complex behavior at the high $f$-values in Fig.\ \ref{poincare diagram} which reveals the existence of a region of P8 LC. The P8 LC is the highest period LC detected in the family of LCs with single large peaks separated by small oscillations. The P8 LC is not shown on the regime map Fig.\ \ref{fc-regime map} due to the narrow size of its region.  
The time series of P6 and P8 LCs in Fig.\ \ref{time series P6 P8} show the characteristic single large peaks separated by small oscillations. The number of sinusoidal source plot periods in each LC in Fig.\ \ref{time series P6 P8} confirms the designations P6 and P8. 

Figure \ref{poincare diagram} shows that the high period LCs all have maximal pulse heights similar to the low-$f$ P1 LC. For $c=0.05$, the amplitude of the maximal pulse is $\gtrapprox 0.9$ out to nearly $f=0.045$, meaning there is not a reduction in the maximal pulse height for these oscillations. The subharmonic family of P1-P4,P6,P8 LCs are superthreshold. (Although for $f>0.045$ the reverse doubling to P2 and P1 LCs consists of subthreshold LCs.) 

Figure \ref{sim-phase-plot} shows numerical $(u,v)$ phase plots for $c=0.05$. The plot on the left for $f=0.0145$ shows the ``filling in" of regions of the phase
\begin{figure}[htbp]
\includegraphics[width=1.6 in]{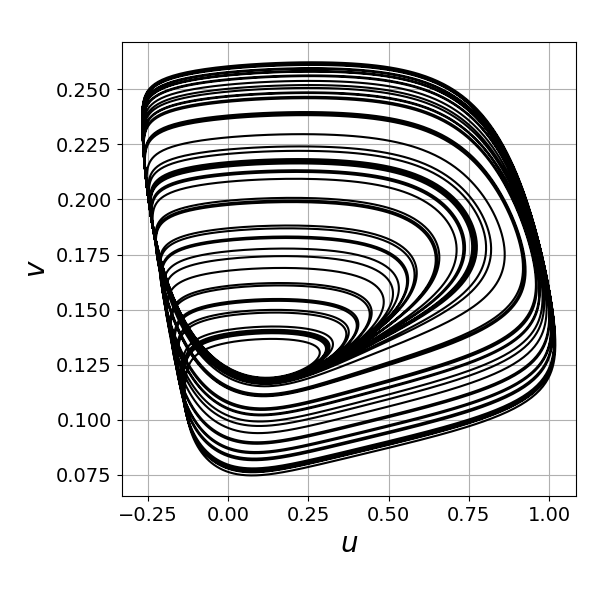}\quad
\includegraphics[width=1.6 in]{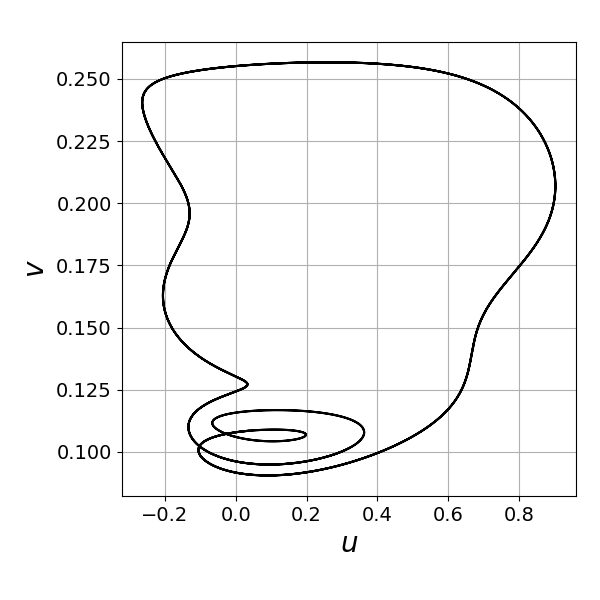}
\caption{Numerical phase plots for $c=0.05$ and: (a) $f=0.0145$ located in the transition region between stable P1 and P2 LCs, (b) $f=0.042$, the P6 LC.}
\label{sim-phase-plot}  
\end{figure}
space that is characteristic of chaotic attractors and is consistent with the Poincar\'e section at $f=0.0145$ in the transition region between stable P1 and P2 LCs seen in Fig.\ \ref{poincare zoom diagram}. Confirmation of chaos by Lyapunov exponents is done the next section. The local maximum values of $u$ in the left phase plot in Fig.\ \ref{sim-phase-plot} range from 0.28 to 1.02 which agree with the Poincar\'e values in Fig.\ \ref{poincare zoom diagram} at $f=0.0145$. The phase plot on the right is for the P6 LC with $f=0.042$ that exists in the P6 regions seen in Figs.\ \ref{f-contin}, \ref{fc-regime map}, \ref{poincare diagram}, and \ref{poincare zoom P8}. This phase plot shows the single large peak and multiple small peaks of $u$ characteristic of the P$n$ family of LCs. 

Next we investigate the behavior for the smaller source amplitude $c=0.02$. The $f$-continuation bifurcation plot in Fig.\ \ref{continuation c=0.02} corresponds to crossing the Fig.\ \ref{fc-regime map} map at $c=0.02$. 
\begin{figure}[htbp]
\includegraphics[width=3 in]{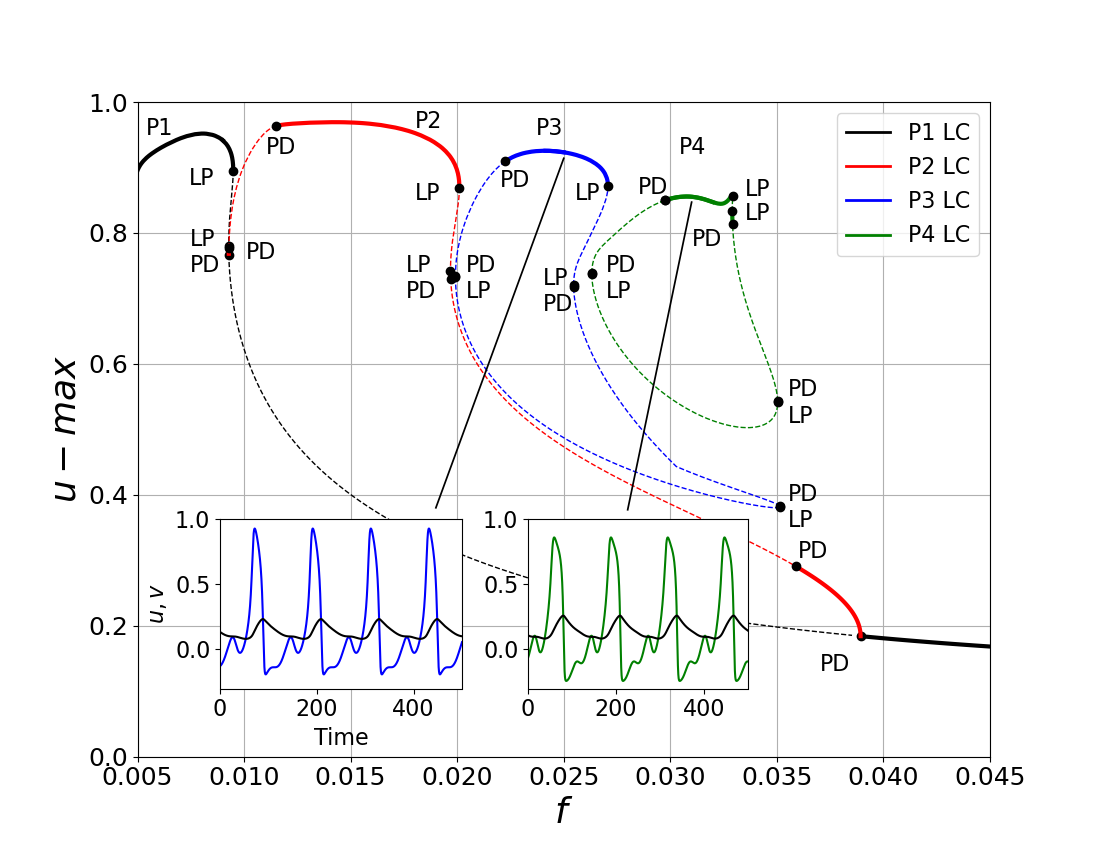}
\caption{Continuation bifurcation diagram for increasing source frequency $f$ for $c=0.02$. Stable (unstable) LC shown as solid (dashed) lines. Insets show stable LC time-series of $u$ and $v$ (black) for P3 at $f=0.025$ and P4 at $f=0.031$.}
\label{continuation c=0.02}  
\end{figure}
The system follows an evolution similar to that for $c=0.05$ in Fig.\ \ref{f-contin} with stable P1 LC ending, then passing through regions of stable higher period P$n$ LCs separated by smaller chaotic regions, and finally regaining stability via reverse-doubling to subthreshold P1 LC at high $f$. (Stable P6 LC was not included in Fig.\ \ref{continuation c=0.02} due to its small size.) 

The insets in Fig.\ \ref{continuation c=0.02} show the time series for the P3 and P4 LCs. Comparison with the P3 and P4 LCs for $c=0.05$ in Figs.\ \ref{f-contin} and \ref{time series P3} shows the same characteristic of single large pulses separated by smaller oscillations. The P$n$ LCs for $n=2\text-4,6,8$ form a family of subharmonic superthreshold LCs of the FHN system with the characteristic of single large pulses separated by small oscillations. Figures \ref{poincare diagram} and \ref{poincare zoom P8} indicate that with the parameters used here, P$n$ LCs with $n=5$ and 7 do not appear in this family. 

The dynamics for the two $c$-values 0.05 and 0.02 do have a notable difference. The difference is the transition route from the stable P1 LC to the large region of stable P2 LC at the low-$f$ side of the island of unstable P1 LC in Fig.\ \ref{fc-regime map}. For $c=0.05$ and increasing $f$, Fig.\ \ref{poincare zoom diagram} indicates that the P1 LC undergoes a period doubling cascade to chaos at the left side of the island. Eventually (after some periodic windows) the chaos reverse-doubles to the stable P2 LC which extends over a large range of $f$ values.  

\begin{figure} [h]
\includegraphics[width=3 in]{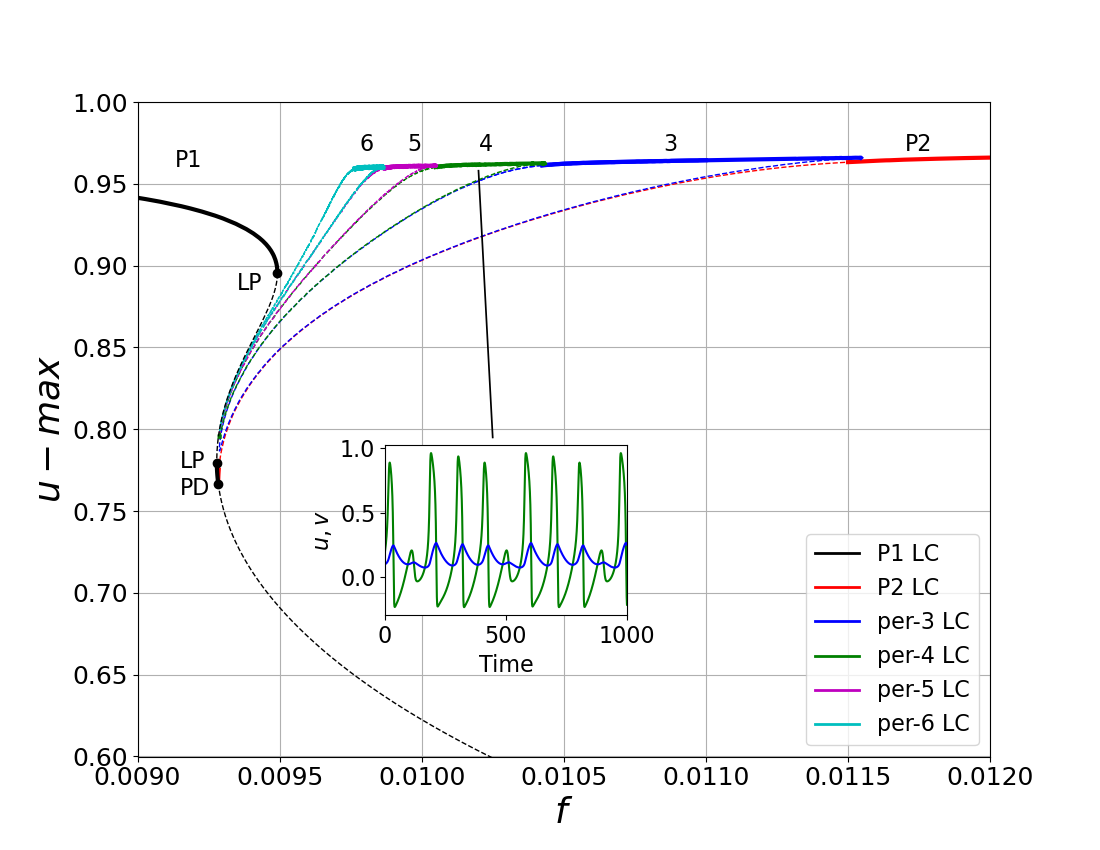}
\caption{Zoomed region showing transition from P1 LC to a series of higher period-$n$ LCs ending in P2 LC for $c=0.02$. Stable (unstable) LC shown as solid (dashed) lines. Inset shows period-4 LC time series at indicated $f$-value.}
\label{continuation c=0.02 zm}  
\end{figure}

For $c=0.02$ a much different transition from stable P1 to stable P2 LC is shown in Fig.\ \ref{continuation c=0.02 zm}, which is a zoom of Fig.\ \ref{continuation c=0.02} showing the transition region, along with some additional higher period LCs. 
For increasing $f$, the P1 LC ends at the LP located at $f=0.0095$, where the system then enters a region of nearly continuous stable LCs. The inset in Fig.\ \ref{continuation c=0.02 zm} shows a time series demonstrating the structure of the LCs in this nearly continuous set. They have single small pulses separated by a larger number of big pulses. As $f$ increases, the LCs have fewer consecutive big pulses, ending in the alternating big and little pulses of P2 LC. LC continuations for $n>6$ are not shown on Fig.\ \ref{continuation c=0.02 zm} due to their narrow ranges, however they are readily found. For example, the period-10 LC (9 large pulses separating single small pulses) is stable from $f=0.009601$ to 0.009621.

Each of the stable LCs in the transition region between P1 and P2 LCs is bounded by a PD on its left and a LP on its right. The boundaries of adjacent stable branches of LCs are close together, giving the appearance in Fig.\ \ref{continuation c=0.02 zm} of continuous stable LCs. Small gaps (not apparent in the figure) between the stable LCs are investigated in the next section using Lyapunov Exponents. 

\begin{figure} [h]
\includegraphics[width=3.4 in]{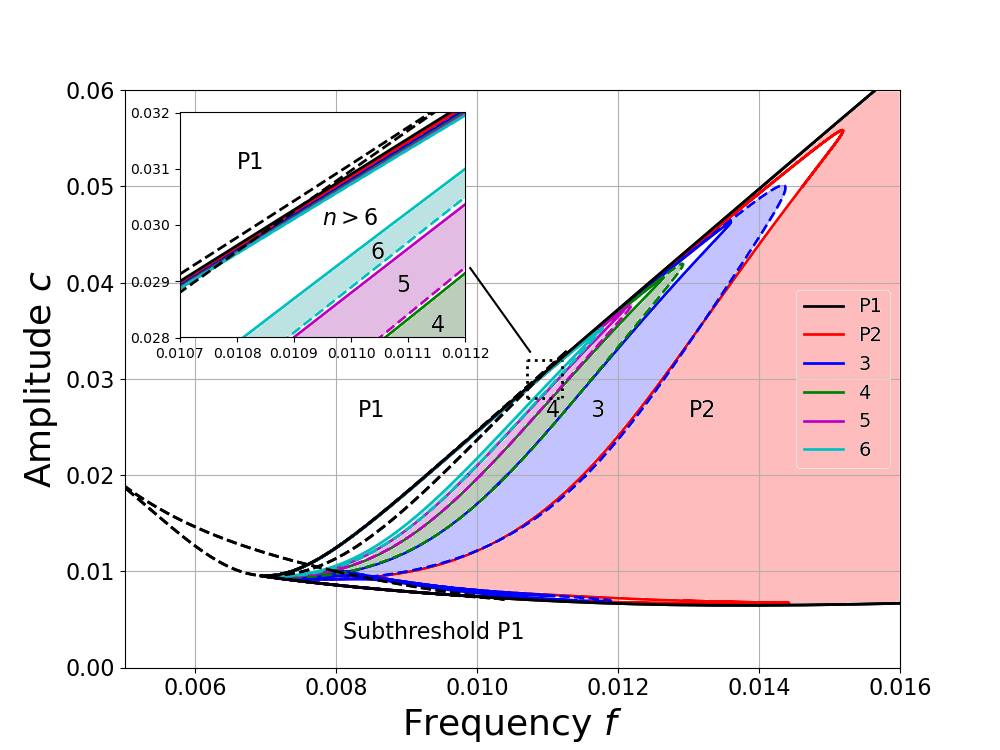}
\caption{Numerical $f$-$c$ regime map showing region of nearly continuous set of period-$n$ LCs. Inset shows zoom of region where the P1 LC's LP and PD lines cross. PD are solid lines, LP are dashed lines.}
\label{fc-regime map zoom}  
\end{figure}

The net effect is that there are two distinct sets of subharmonic LCs. The set with single large pulses separated by multiple small oscillations, designated the P$n$ set, has stable LCs throughout much of the island as seen in Fig.\ \ref{fc-regime map}. The other set, referred to here simply as the period-$n$ LCs, are the tightly packed LCs characterized by multiple large pulses separated by single small pulses. This family of period-$n$ LCs exists in the transition region shown in Fig.\ \ref{fc-regime map zoom} between the P1 and P2 LCs. This region is located at the low $f$ and low $c$ portion of the island in Fig.\ \ref{fc-regime map}. 
 
The P2 LC is a member of both families since it consists of single large pulses separated by single small pulses. The P1 LC can be either subthreshold or superthreshold. At sufficiently small values of $c$, the threshold for positive feedback in Eq.\ \eqref{nonauton-eqns} is not met resulting in subthreshold P1 oscillations. These are in the region labeled ``Subthreshold P1" in Fig.\ \ref{fc-regime map zoom}. Subthreshold P1 also occur at larger values of $c$ where the frequency is too high, as evidenced by the amplitude of the high-$f$ P1 LC in the $f$-continuation diagrams in Figs.\ \ref{f-contin} and \ref{continuation c=0.02}.  

The nature of the transition from P1 LC to P2 LC is determined by its location in the parameter space of Fig.\ \ref{fc-regime map} relative to the narrow region of tightly packed period-$n$ LCs featured in Fig.\ \ref{fc-regime map zoom}. Figure \ref{fc-regime map zoom} shows that the region of period-3 LC stops just beyond $c=0.05$ (blue dashed line), and period doubling beyond P2 LC (i.e. P4,P8...) stops at $c=0.056$ (red line). Thus, for $c>0.056$ the transition is a simple period-doubling of P1 to P2 LC. For the smaller value $c=0.05$ Fig.\ \ref{poincare zoom diagram} shows that the P1 PD is the beginning of a period-doubling cascade to chaos followed by reverse doubling to the P2 LC, with a window of the period-3 LC inside the chaos. 

\begin{figure} [h]
\includegraphics[width=3 in]{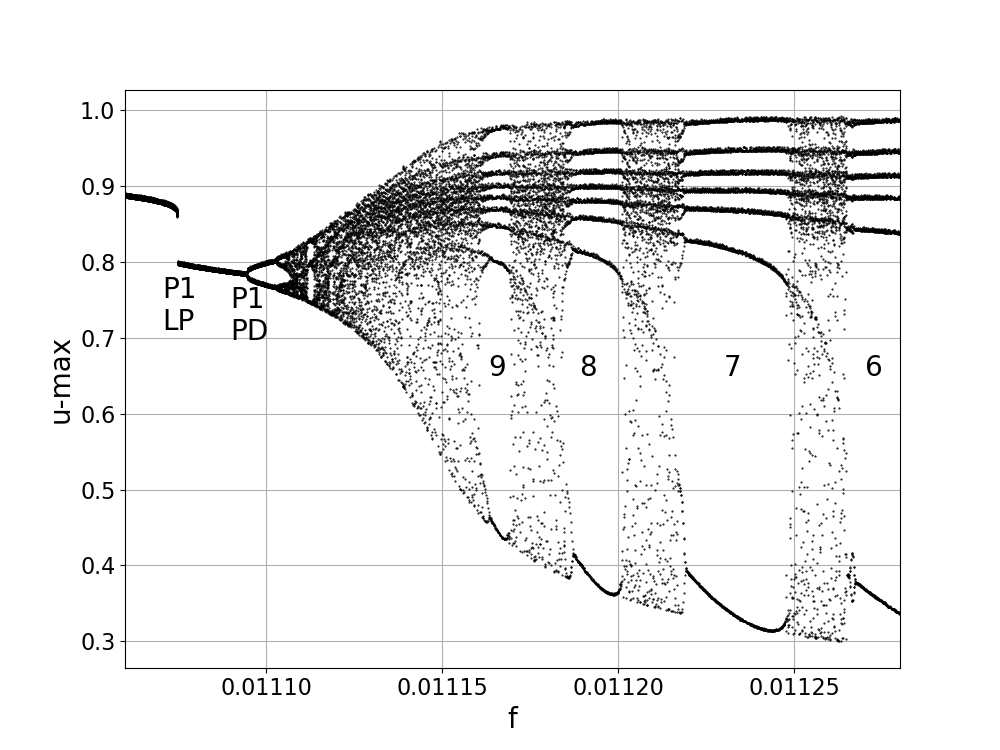}
\caption{Poincar\'e plot for transition from P1 LC to the period-$n$ LC sequence between P1 and P2 LCs at $c=0.0315$. }
\label{poincare diagram c=0.0315}  
\end{figure}

Now we consider what happens to the P1 to P2 LC transition with further decrease of $c$. The inset in Fig.\ \ref{fc-regime map zoom} shows a zoom of the border region between P1 LC and the sequence of tightly packed period-$n$ LCs. 
The P1 PD line (solid black) intersects one of the P1 LP lines (dashed black) at ($f=0.01092,c=0.0304)$, meaning that for an increasing $f$ the stable P1 LC ends at a PD when $c>0.0304$ and ends at an LP when $c<0.0304$. Figure \ref{poincare diagram c=0.0315} shows a Poincar\'e bifurcation plot for $c=0.0315$, where the two LPs bounding the region of P1 hysteresis occur at lower $f$-values than the P1 PD. 
The labeled P1 LP at $f=0.011075$ is where the region of P1 LC hysteresis ends causing the switch to the lower P1 LC. The P1 PD at $f=0.011095$ initiates a period-doubling cascade to chaos, which extends to the appearance of period-9 LC followed by the sequential reduction of period-$n$ LCs. The period-9 through period-7 LCs in Fig.\ \ref{poincare diagram c=0.0315} exist in the region labeled $n>6$ in Fig.\ \ref{fc-regime map zoom}, thus indicating the continued sequential reduction to the P2 LC at $f=0.0128$.    
For $c<0.0304$ Fig.\ \ref{fc-regime map zoom} shows that one of the P1 LPs is at a higher $f$-value than the P1 PD (since the solid black line is between the two dashed black lines). An example is the P1 LP at $f=0.0095$ in Fig.\ \ref{continuation c=0.02 zm} for $c=0.02$. As $f$ increases, the P1 LC switches directly to the set of period-$n$ LCs instead of period doubling. 

\begin{figure} [h]
\includegraphics[width=1.6 in]{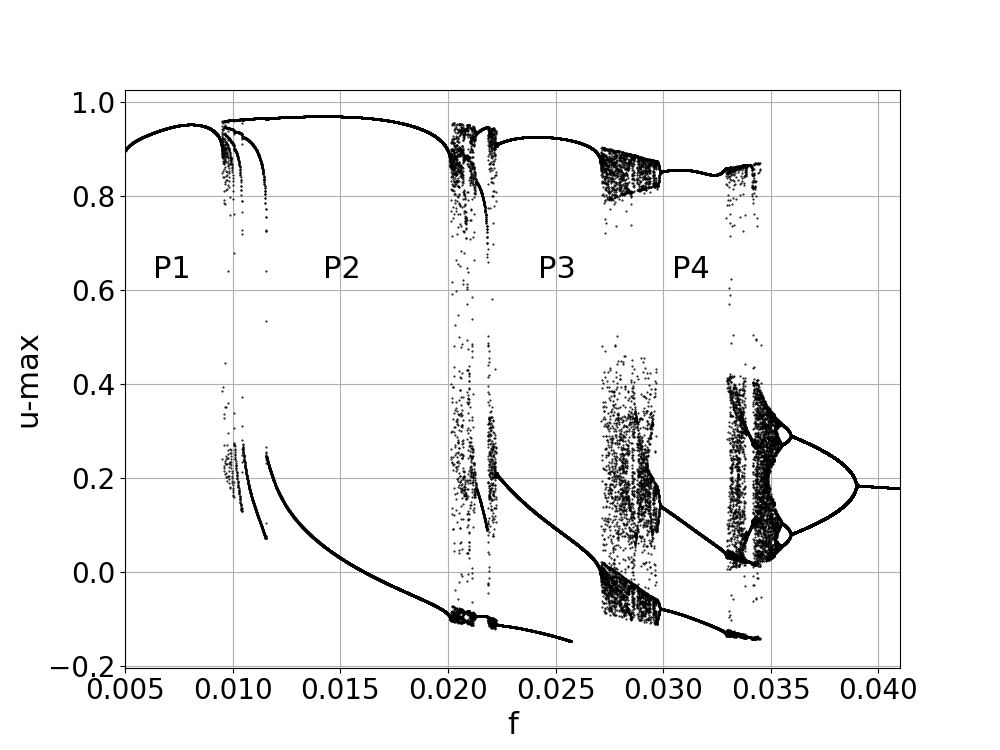}\quad
\includegraphics[width=1.6 in]{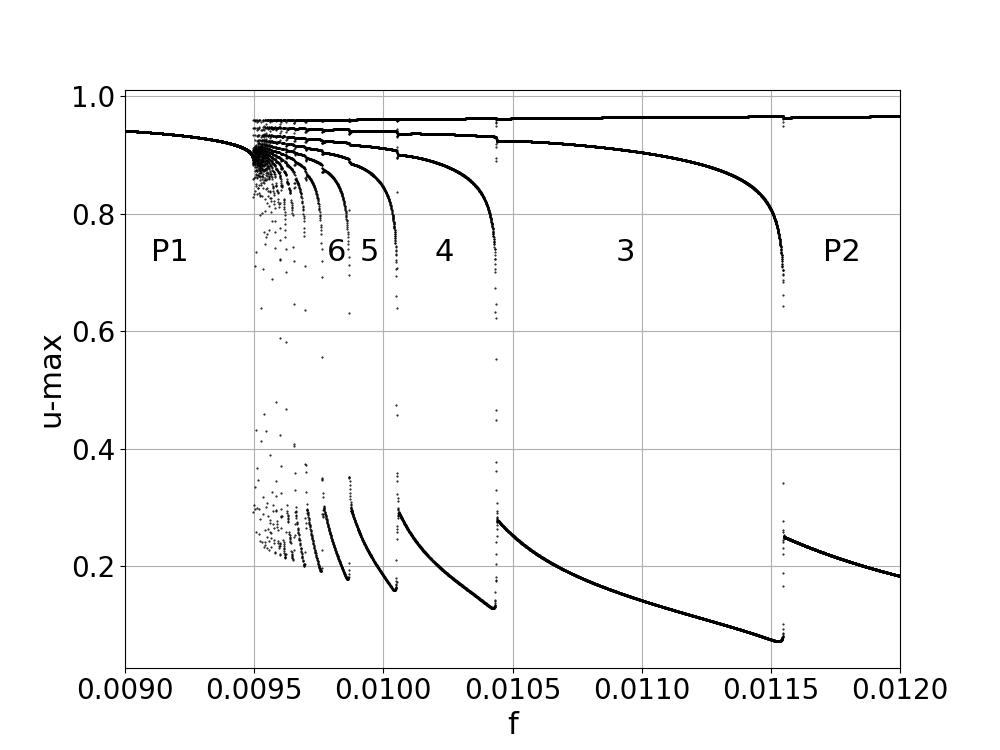}
\caption{Left: Poincar\'e bifurcation diagram for increasing source frequency $f$ for $c=0.02$. Right: Zoomed region showing transition from P1 LC to a decreasing series of period-$n$ LCs ending in P2 LC. }
\label{poincare diagram c=0.02}  
\end{figure}

Figure \ref{poincare diagram c=0.02} shows the Poincar\'e bifurcation plot for $c=0.02$ over the entire range of unstable P1 LC. The left panel shows the regions of stable P1, P2, P3, and P4 LCs consistent with the continuation plot Fig.\ \ref{continuation c=0.02}.   
The right panel shows a zoom of the transition from stable P1 LC, through the set of nearly continuous stable period-$n$ LCs ending in stable P2 LC. The sequence of tightly packed period-$n$ LCs in the transition region is consistent with its appearance in the continuation bifurcation in Fig.\ \ref{continuation c=0.02 zm}. 

\begin{figure} [h]
\includegraphics[width=1.6 in]{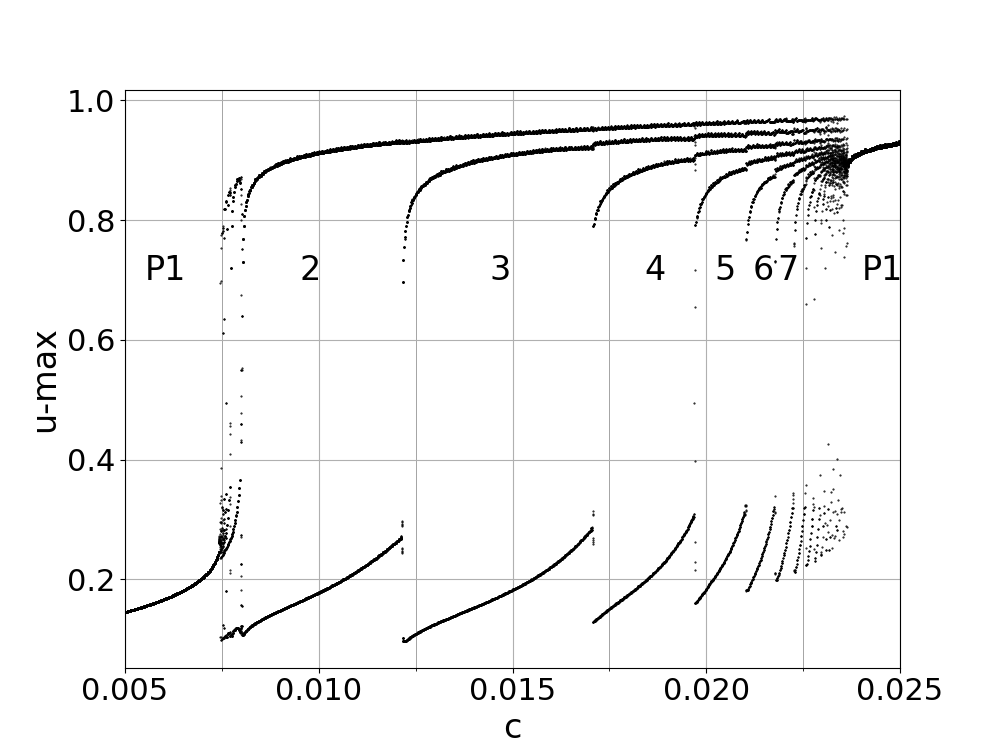}\quad
\includegraphics[width=1.6 in]{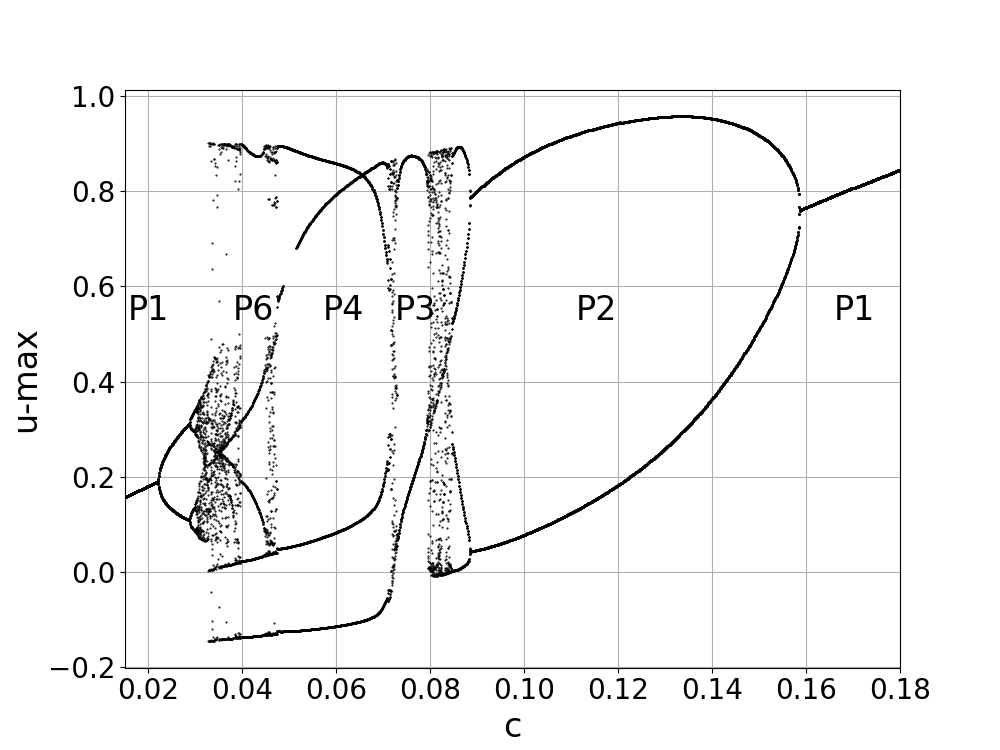}
\caption{Poincar\'e bifurcation diagrams for increasing source amplitude $c$ for: Left $f=0.01$, Right $f=0.04$.}
\label{poincare-c diagrams}  
\end{figure}

It is also of interest to consider traversing the regime map Fig.\ \ref{fc-regime map} over $c$ values at fixed frequency. Figure \ref{poincare-c diagrams} shows Poincar\'e bifurcation plots for increasing $c$ at fixed frequencies $f=0.01$ and 0.04. 
The left panel for $f=0.01$ shows the system starting in subthreshold P1 LC, then progressing through the series of period-$n$ LCs to the P1 LP at $c=0.0236$ where the system goes to superthreshold P1 LC. The right panel for $f=0.04$ shows the system starting in subthreshold P1 LC, then period doubling to chaos before entering the family of P$n$ LCs separated by chaotic regions, and finally reverse doubling to P1 LC. Both of these bifurcation plots are consistent with predictions based on corresponding traverses of Fig.\ \ref{fc-regime map} at $f=0.01$ and 0.04. In particular, it is clear that only the period-$n$ LCs will be encountered at $f=0.01$ and only the P$n$ LCs at $f=0.04$. 

\subsection{Lyapunov Exponents}
LE calculations detect parameter ranges with chaotic behavior and are therefore complementary to the parameter regime maps created by bifurcation analysis since those methods do not detect chaos. In LE plots produced from autonomous systems, stable LCs are indicated by the maximal LE being zero, whereas chaotic behavior has a positive maximal LE. 

Information about PD points can also be indicated on the LE plot. When a changing parameter causes a stable LC to undergo period doubling, the maximal LE remains at zero throughout the process. However, the second maximal LE increases until it reaches zero at the PD bifurcation point, then it returns to increasingly negative numbers.

\begin{figure} [h]
\includegraphics[width=3 in]{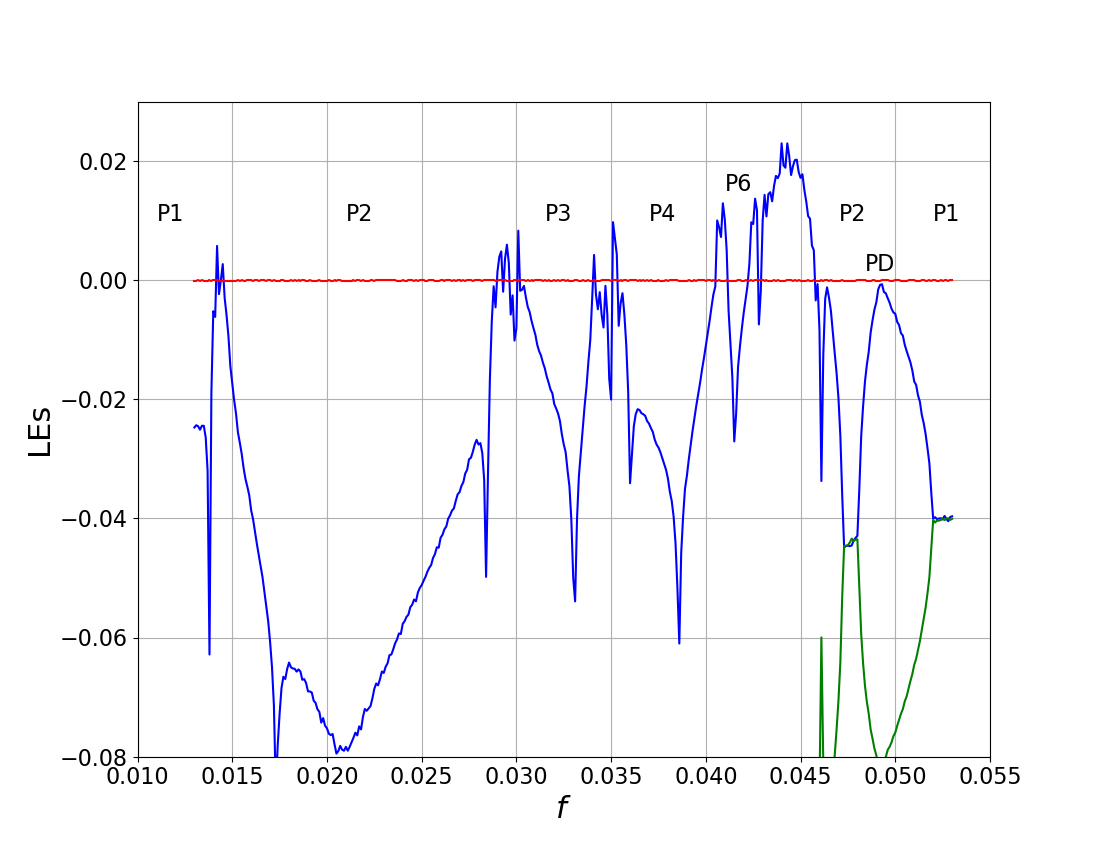}
\caption{LE plot versus source frequency for FHN system with source amplitude $c=0.05$. Chaos occurs where the blue LE is positive. Regions of stable LC are indicated.}
\label{LE}  
\end{figure}

The LE plot in Fig.\  \ref{LE} for source amplitude $c=0.05$ shows the stable LCs P1, P2, P3, P4, and P6 occurring where the red LE is zero and maximal. In between these stable LCs are regions where the blue LE is maximal and positive indicating chaos. 
Smaller periodic windows of stable LC can also exist within the the chaotic regions. The reverse doubling back to P1 LC at high $f$ which is clearly seen in Fig.\ \ref{poincare diagram} is also apparent in Fig.\ \ref{LE} by the labeled PD point where the second maximal LE (blue) reaches zero.    

\begin{figure} [h]
\includegraphics[width=3 in]{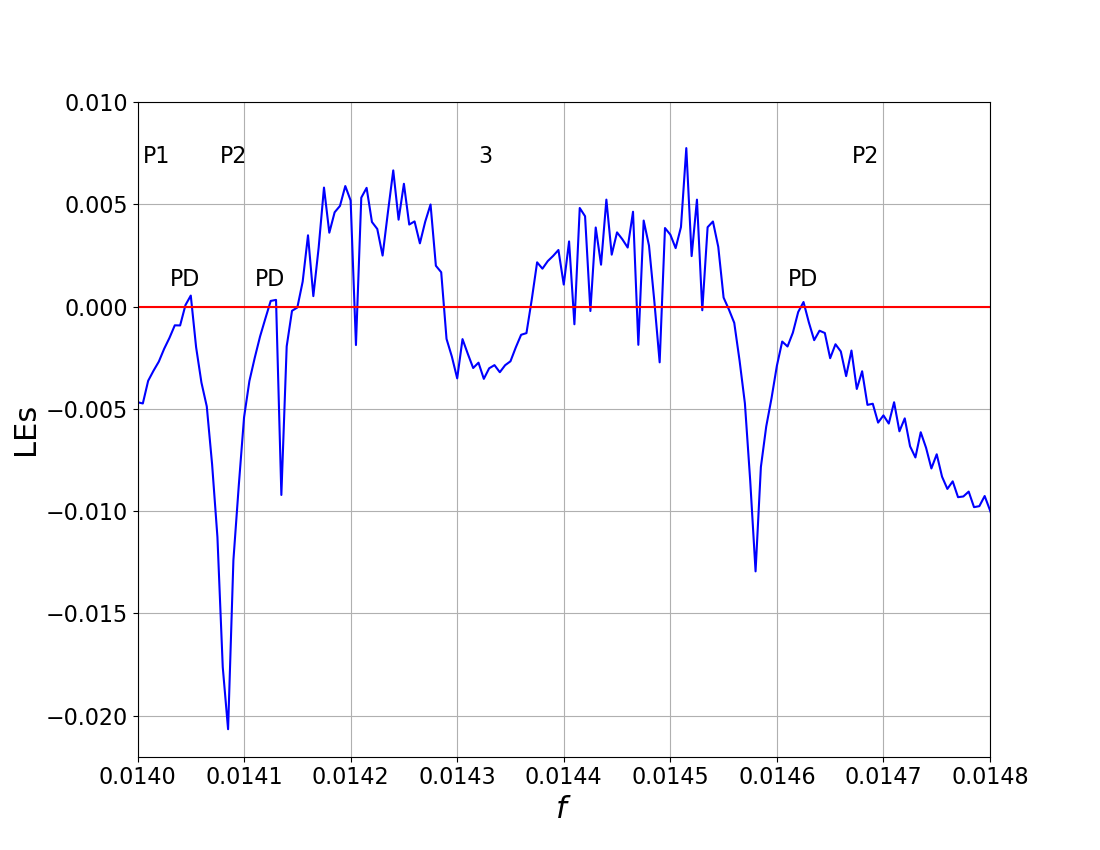}
\caption{Zoom of LE plot versus source frequency for FHN system with source amplitude $c=0.05$ showing the transition from P1 through chaos to P2 LC. A window of the period-3 LC appears in the chaos.}
\label{LE-zoom}  
\end{figure}

Figure \ref{LE-zoom} is a zoom of the LE plot showing the transition from P1 LC to P2 LC depicted in the Poincar\'e plot Fig.\ \ref{poincare zoom diagram}. 
The first two PD points of the cascade to chaos are indicated on Fig.\ \ref{LE-zoom} as is the PD for the reverse doubling to the P2 LC. These figures demonstrate that for $c=0.05$ the boundaries of the stable P1 and P2 LCs in this transition region are period doubling cascades to chaos. The figures also show the appearance of a single member of the period-$n$ LC family, the period-3, within the chaos.   

\begin{figure} [h]
\includegraphics[width=3 in]{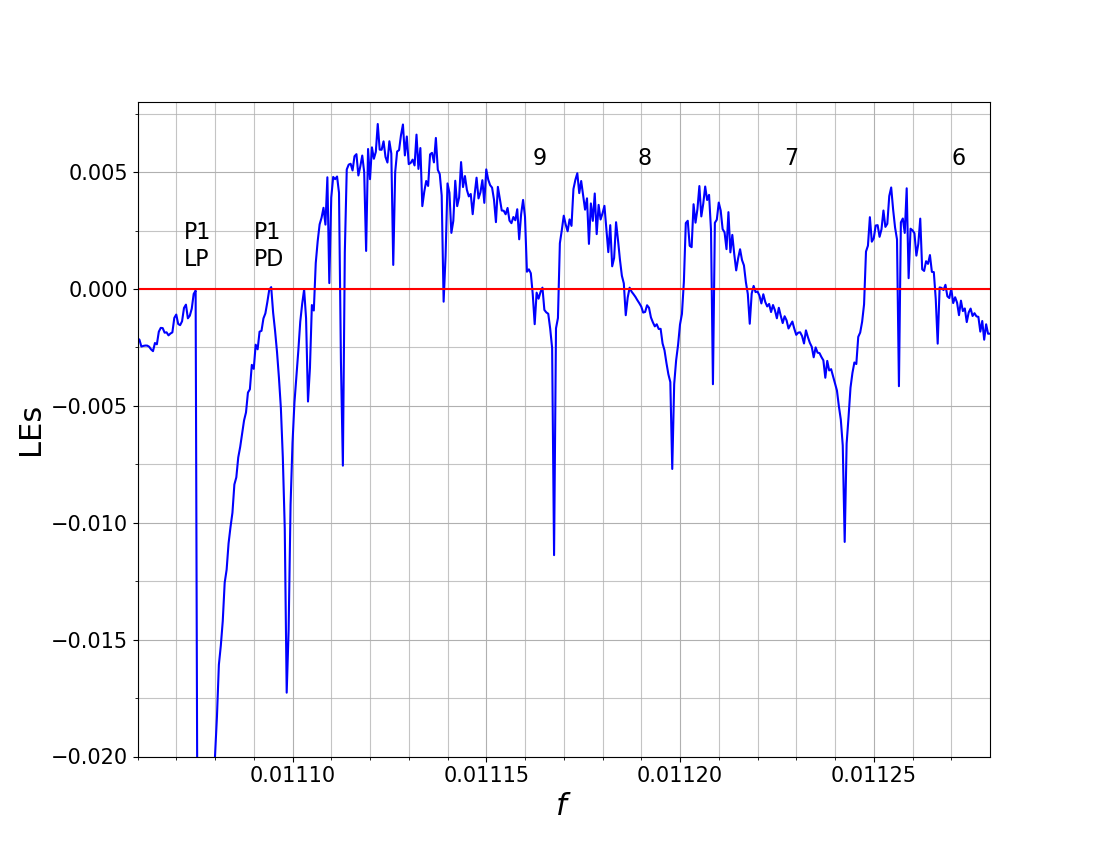}
\caption{LE plot of the transition from P1 LC to the period-$n$ LC sequence between P1 and P2 LCs for $c=0.0315$.}
\label{LE c=0.0315}  
\end{figure}

Now consider the P1 to P2 LC transition region for $c=0.0315$ where Fig.\ \ref{poincare diagram c=0.0315} shows a period doubling cascade to chaos and the appearance of multiple period-$n$ LCs. The LE plot in Fig.\ \ref{LE c=0.0315} shows the period-doubling cascade to chaos initiated at P1 PD and 
the chaos in the gaps between the stable period-$n$ LCs. The end of the P1 LC hysteresis at P1 LP is also apparent, consistent with Fig.\ \ref{poincare diagram c=0.0315}.

\begin{figure} [htbp]
\includegraphics[width=3 in]{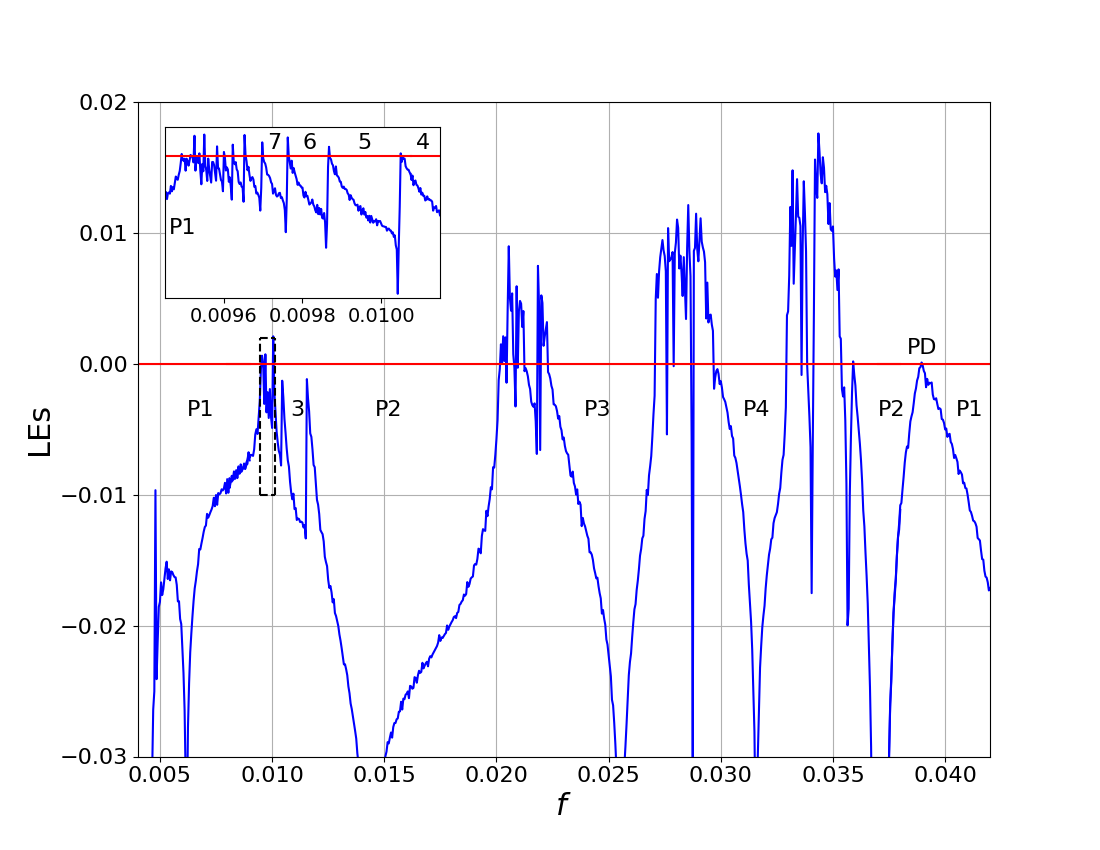}
\caption{The two maximal LEs versus source frequency for FHN system with source amplitude $c=0.02$. Inset shows zoom of region inside dashed lines. }
\label{LE c=0.02}  
\end{figure}

Figure \ref{LE c=0.02} shows the LE plot for the smaller source amplitude $c=0.02$. The plot is similar to Fig.\ \ref{LE}, showing the progression from P1 LC through the higher period P$n$ LCs separated by ranges of chaos, then reverse doubling back to P1 LC. 
A notable difference occurs in the transition region between the P1 and P2 LCs at the low $f$ values. The inset of Fig.\ \ref{LE c=0.02} shows a zoom of the dashed line region of the main graph. The range of nearly continuous stable period-$n$ LCs in Fig.\ \ref{continuation c=0.02 zm} is seen in Fig.\ \ref{LE c=0.02}, beginning with the P2 and period-3 LCs in the main graph, then the period-4 through period-7 LCs in the inset. The zoomed inset shows narrow positive LE spikes in the transitions between stable LCs, suggesting narrow windows of chaos.

\begin{figure} [htbp]
\includegraphics[width=3 in]{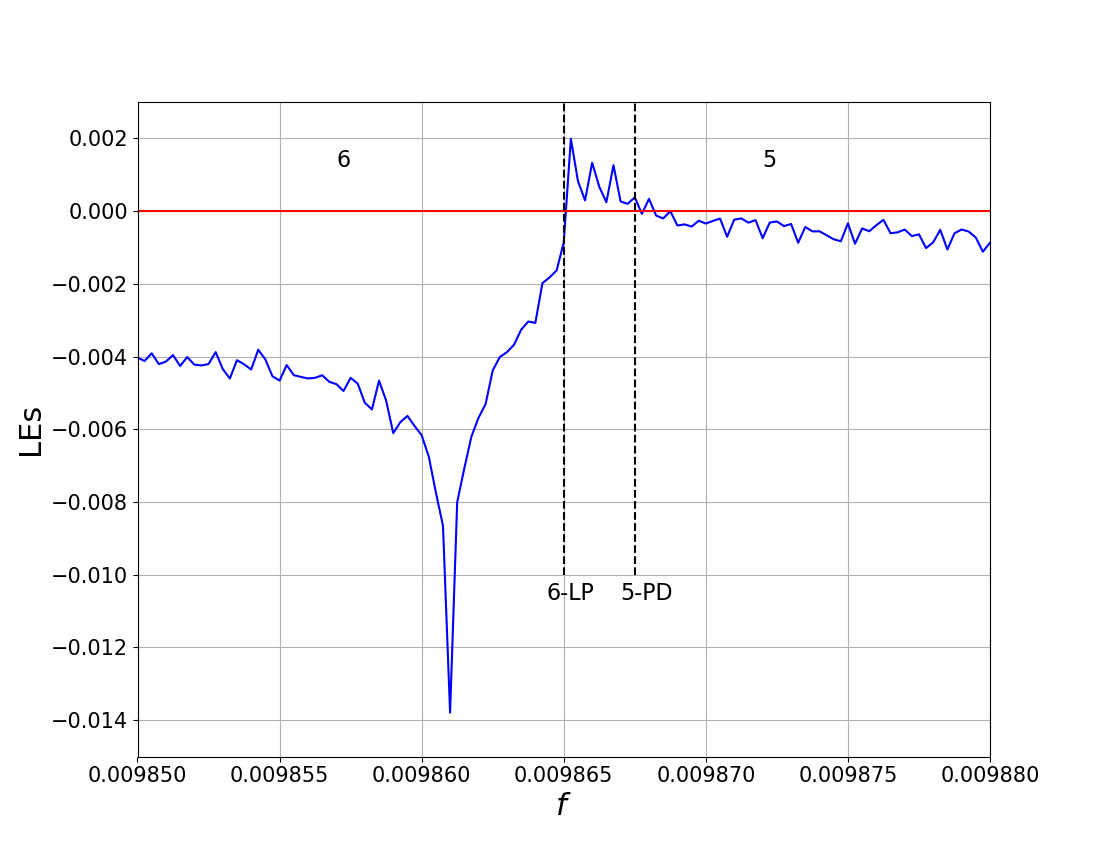}
\caption{The two maximal LEs versus source frequency for the transition from period-5 to 6 LC. $c=0.02$.}
\label{LE 5 to 6}  
\end{figure}

One of these narrow positive LE spikes was investigated with higher resolution. Figure \ref{LE 5 to 6} shows a zoomed LE plot of the transition region between period-5 and period-6 LCs for $c=0.02$. 
The gap between stable portions of these LCs is the region between the period-6 LP and period-5 PD. Inside the gap the blue LE is maximal, positive, and relatively small, indicating weak chaos. Outside of this region the red LE is maximal and zero, consistent with stable LCs. 

The narrow regions of chaos in the zoom inset in Fig.\ \ref{LE c=0.02} are due to gaps like that in Fig.\ \ref{LE 5 to 6} between stable period-$n$ and $n+1$ LCs for $n\ge 5$. For $n<5$ and $c=0.02$, the gap disappears and can result in multistability. For example, the period-3 and 4 LCs are both stable between $f=0.010425$ and 0.10433, consistent with the maximal LE=0 (red) over this range and the $2^{nd}$-maximal LE (blue) being negative in the main graph of Fig.\ \ref{LE c=0.02}. 

\subsection{Circuit Measurements}
\begin{figure} [h]
\includegraphics[width=1.6 in]{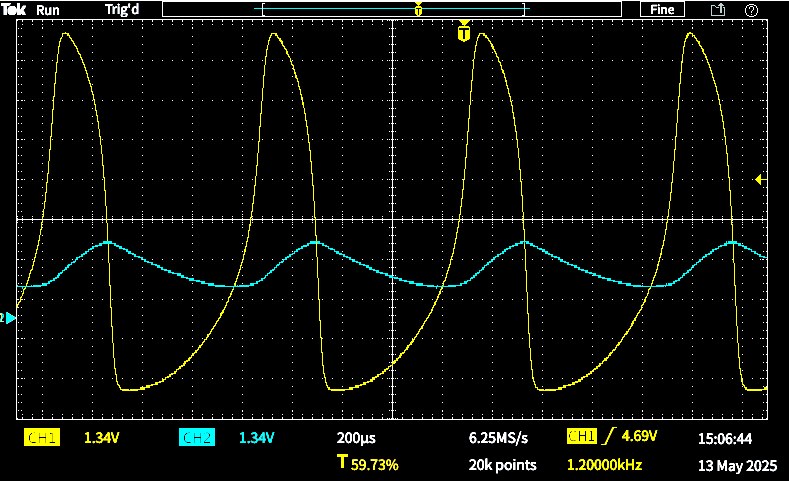}\quad 
\includegraphics[width=1.6 in]{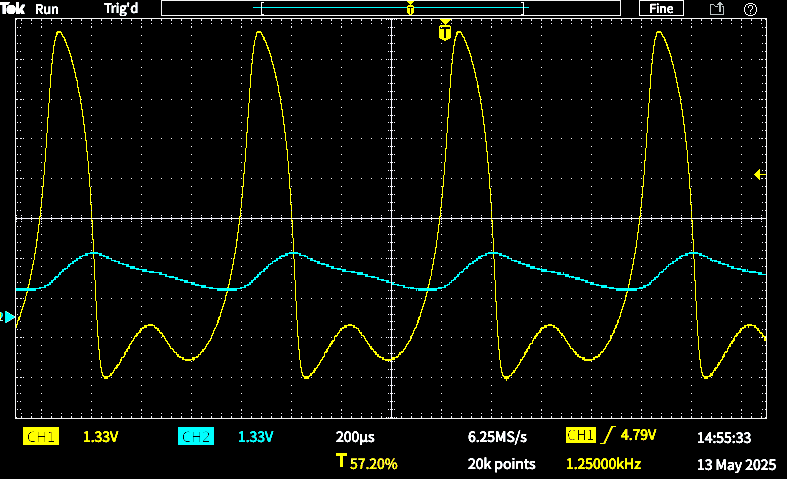}\\
\includegraphics[width=1.6 in]{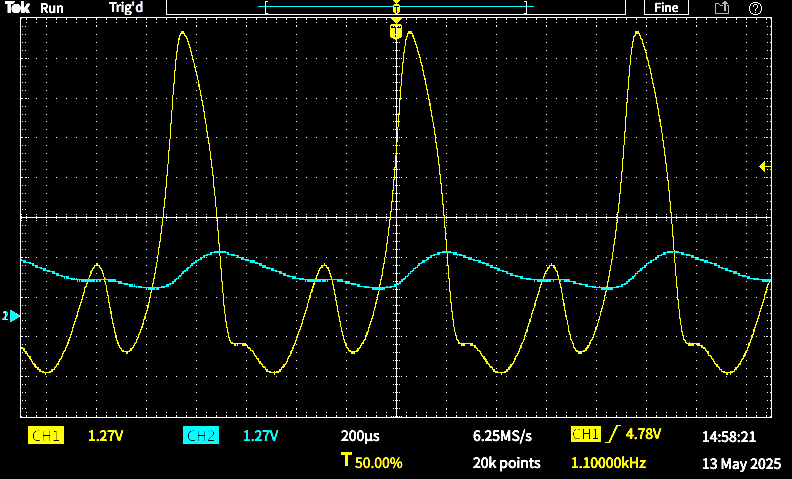}\quad 
\includegraphics[width=1.6 in]{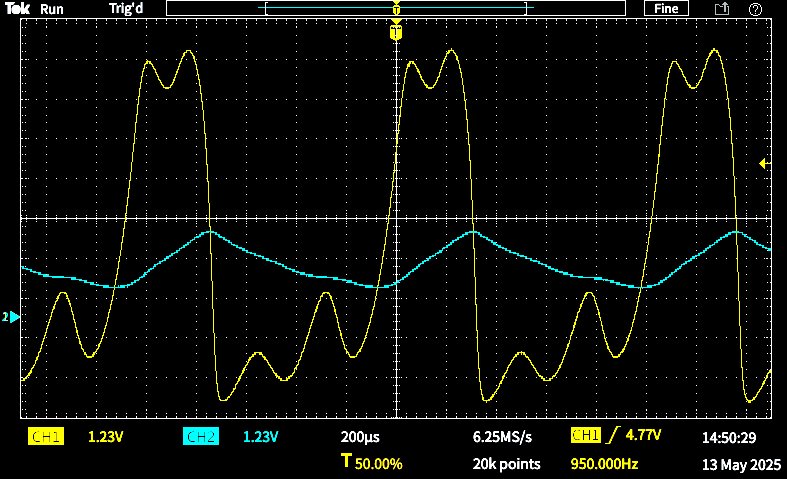}\\
\includegraphics[width=1.6 in]{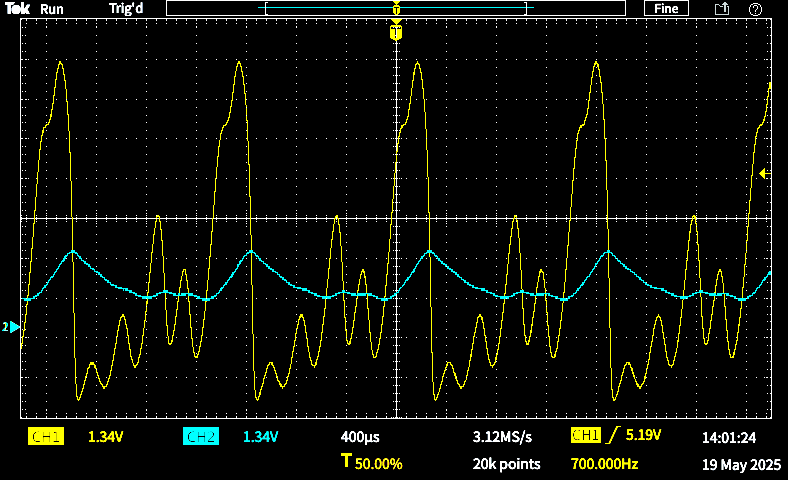}\quad 
\includegraphics[width=1.6 in]{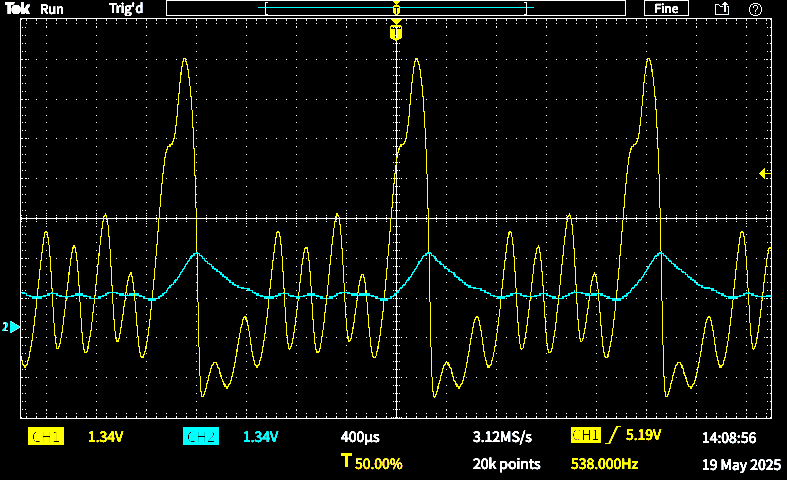}
\caption{Oscilloscope screenshots of time series for the circuit voltages $V_u$ (yellow) and $V_v$ (blue). Sinusoidal source amplitude is 0.5 volts. Frequency is given for each LC. P1 (1.2 kHz) and P2 (2.5 kHz) top row. P3 (3.3 kHz) and P4 (3.8 kHz) middle row. P6 (4.2 kHz) and P8 (4.31 kHz) bottom row. }
\label{screenshots}  
\end{figure}

Oscilloscope screenshots of the time-series of the voltages $V_u$ and $V_v$ measured in the circuit Fig.\ \ref{circuit} are shown in Fig.\ \ref{screenshots}. The LCs are the P1, P2, P3, P4, P6, and P8. 
The numerical LC time-series for P1, P2, and P4 seen in the three insets of Fig.\ \ref{f-contin} and for P3, P6, and P8 LCs in Figs.\ \ref{time series P3} and \ref{time series P6 P8} all show good agreement with their corresponding measured voltages in Fig.\ \ref{screenshots}. The circuit measurements were done using sinusoidal source amplitude corresponding to the $c=0.05$ used in the numerical simulations. The source frequencies agree well with the dimensionless $f$ values. For example, the 1.2 kHz for P1 in Fig.\ \ref{screenshots} gives the $f=0.012$ used for the P1 LC inset of Fig.\ \ref{f-contin}. The 4.31 kHz used for P8 in Fig.\ \ref{screenshots} agrees well with $f=0.0428$ in Fig.\ \ref{time series P6 P8}. It is unrealistic to expect perfect agreement. 

The term ``amplitude" is ambiguous. In the world of electronics ``amplitude" often refers to the peak-to-peak value of a sinusoidal signal. Here, amplitude of the sinusoid refers to the coefficient multiplying the sinusoidal function. Applying a 1 volt peak-to-peak sinusoidal voltage corresponds to 0.5 volt amplitude and dimensionless amplitude $c=0.05$. 

\begin{figure} [h]
\includegraphics[width=3.2 in]{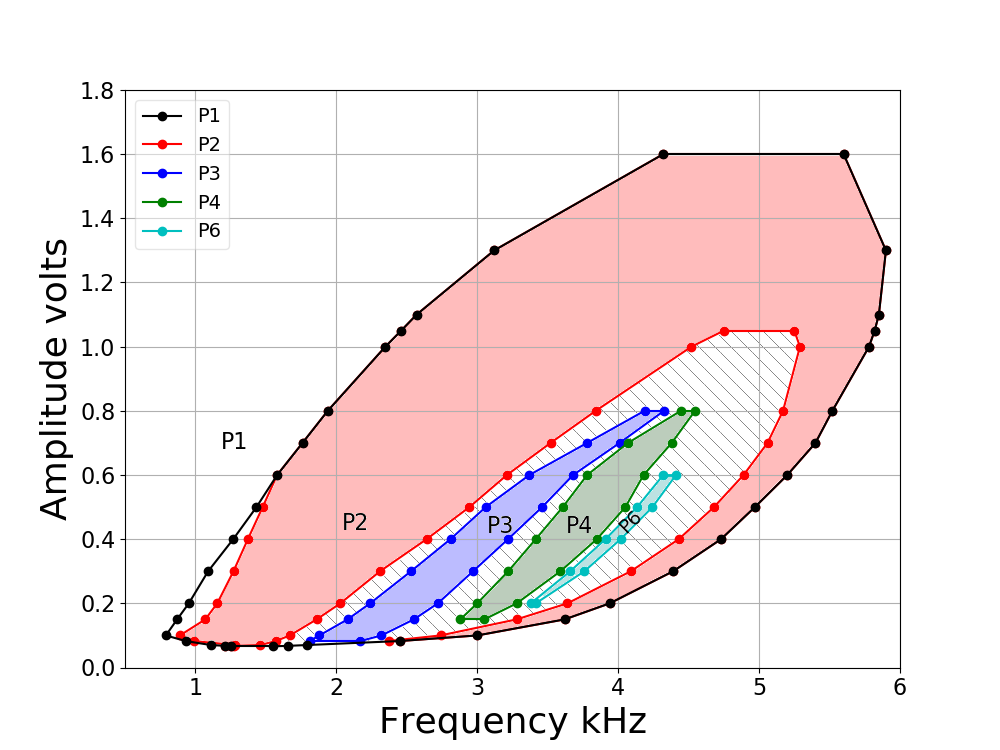}
\caption{Experimental regime map measured from the sinusoidally driven FHN circuit showing regions of P1-P4 and P6 LCs. Circles are data points. Chaos indicated by diagonal lines.}
\label{circuit map}  
\end{figure}

Figure \ref{circuit map} shows the $f$-$c$ regime map measured from the circuit. 
The borders of the regimes in the map are easily found by watching the oscilloscope trace while varying the frequency. The change from stable LC to unstable behavior is especially clear when the oscilloscope is set in x-y mode to display the $V_u$-$V_v$ phase plot. The strong resemblance of the maps for numerical predictions and circuit measurements in Figs.\ \ref{fc-regime map} and \ref{circuit map} is clear. The same regions of the stable P1 through P6 LCs are found in both maps.   

\begin{figure} [h]
\includegraphics[width=1.6 in]{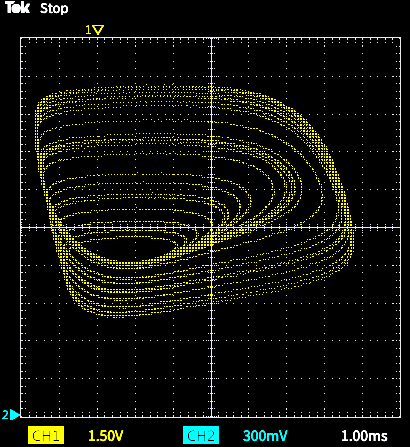}\quad
\includegraphics[width=1.6 in]{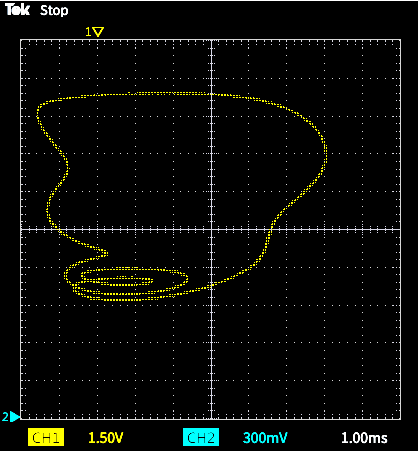}
\caption{Oscilloscope screenshots of phase plots of measured voltages corresponding to variables $u$ and $v$ for (a) chaos ($c=0.05, f=0.0144$) and (b) P6 LC ($c=0.05, f=0.0419$).} 
\label{circuit-phase-plots}  
\end{figure} 
The oscilloscope phase plot in the left panel of Fig.\ \ref{circuit-phase-plots} shows the chaotic circuit behavior using a 1 volt peak-to-peak ($c=0.05$), 1.44 kHz ($f=0.0144$) sinusoidal source. This amplitude and frequency correspond to the numerical prediction in Fig.\ \ref{sim-phase-plot} for the transition region from stable P1 to stable P2 LC. The visual resemblance of the left panels of Figs.\ \ref{sim-phase-plot} and \ref{circuit-phase-plots} is clear. 

The right panel in Fig.\ \ref{circuit-phase-plots} shows the phase plot measured from the circuit using a 1 volt peak-to-peak 4.19 kHz source. It agrees well with the P6 LC numerical simulation in the right panel of Fig.\ \ref{sim-phase-plot} using $c=0.05$ and $f=0.042$. 

\begin{figure} [h]
\includegraphics[width=1.6 in]{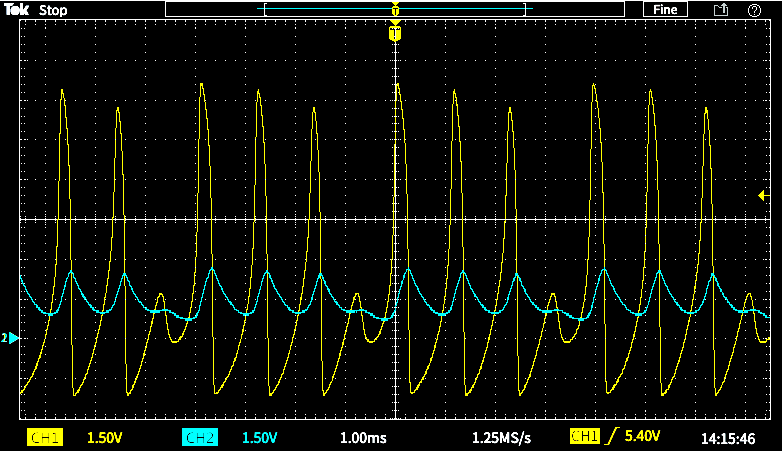}\quad
\includegraphics[width=1.6 in]{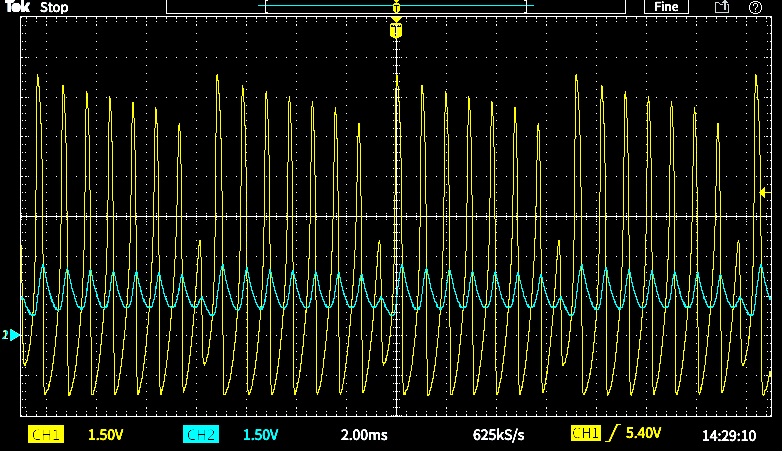}
\caption{Oscilloscope screenshots of time series of measured voltages $V_u$ (yellow) and $V_v$ for (a) period-4 LC ($c=0.02, f=0.0102$) and (b) period-8 LC ($c=0.0315, f=0.01114$).} 
\label{per-n TS}  
\end{figure}

Now we consider circuit measurements of the family of closely packed period-$n$ LCs in the transition region between the stable P1 and P2 LCs for small $f$ and small $c$. It is impractical to experimentally measure the regimes for these LCs due to the large number of LCs and the small parameter ranges in which they exist. Instead we compare two cases of measured and simulated time-series. Figure \ref{per-n TS} shows measurements of the period-4 and period-8 LCs, both clearly displaying the characteristic of
single small pulses separated by multiple large pulses. The left panel in Fig.\ \ref{per-n TS} shows the measured period-4 LC using 0.4 volt peak-to-peak 1.02 kHz source, corresponding to $c=0.02$, $f=0.0102$ used for the period-4 LC simulation in the inset of Fig.\ \ref{continuation c=0.02 zm}. The right panel in Fig.\ \ref{per-n TS} shows the measured period-8 LC using 630 mV peak-to-peak 1.114 kHz source. This graph agrees well with Fig.\ \ref{poincare diagram c=0.0315} for $c=0.0315$, which predicts the region of period-8 LC located for $f$ in 0.01118 to 0.0112. The small discrepancy between the 1.114 kHz source and the $f$-range is not significant, considering the fine resolution of these values and the tolerance of the electrical components.  

\section{Discussion}
The focus of this work is the application of sinusoidal forcing to an otherwise at rest FHN excitable system. The first goal is to present the dynamical behavior in a 2d regime map in the parameter space of source frequency and amplitude. The other major goal is to demonstrate good agreement between the results from numerical simulations and from circuit measurements. The purpose of the analog circuit measurements is to establish validity of the numerical simulations. 

It is natural to question what purpose is being served by pursuing these goals for such a well-studied system. The fact that studies with similar goals were done long ago means that those studies did not benefit from the computational capabilities and bifurcation analysis methods available today which allow a more comprehensive analysis and presentation of the dynamical behavior. The intention is to present a clearer and more thorough picture than currently exists of the rich variety of behavior demonstrated by this relatively simple system. Providing this deeper insight could be beneficial to the widespread ongoing research which makes use of the FHN system. 

We begin by considering the numerical simulation and circuit measurement regime maps, Figs.\ \ref{fc-regime map} and \ref{circuit map}, and noting their strong resemblance. The region of complex behavior forms an island in a sea of stable limit cycle with the source frequency. Source frequencies which are too small or too large result in oscillations at the same frequency as the source. Source frequencies near the system's natural frequency ($\approx \epsilon=0.01$) can result in subharmonic LCs and complex behavior, thereby demonstrating resonance response. Amplitudes which are too small ($c\lesssim 0.007$ for the constant $i=0.1$ used here) are subthreshold and are therefore able to induce only small amplitude P1 oscillations over the entire frequency range. Amplitudes that are too big ($c>0.18$) overwhelm the FHN system resulting in large amplitude P1 LC. The border of the island is predominantly the period-doubling (PD) between stable P1 and stable P2 LC.  
 
Two distinct families of subharmonic LCs exist in the island of unstable P1 LC. The regime maps show that a large portion of the island consists of stable P$n$ LCs with $n>1$, characterized by single large pulses separated by smaller pulses, exemplified by Figs.\ \ref{f-contin}, \ref{time series P3}, \ref{time series P6 P8}, \ref{continuation c=0.02}, and \ref{screenshots}. There are gaps of chaotic behavior between these LCs seen in the Poincar\'e plots Figs.\ \ref{poincare diagram}, \ref{poincare zoom P8}, and \ref{poincare diagram c=0.02}. Interestingly, $n$ takes on only the values (2-4,6,8) in the island for the parameter values used in this study. 

The other family of subharmonic LCs is the period-$n$, characterized by single small pulses separated by multiple large pulses, exemplified by Figs.\ \ref{continuation c=0.02 zm} and \ref{per-n TS}. This family exists in a small region of the island featured in Fig.\ \ref{fc-regime map zoom} at low $f$ and low $c$ values in the transition region between stable P1 and P2 LCs. The period-$n$ LCs can be adjacent to one another as in Fig.\ \ref{poincare diagram c=0.02} and \ref{poincare-c diagrams}, or they can be separated by chaotic gaps as in Fig.\ \ref{poincare diagram c=0.0315}. 

The period-$n$ LCs are restricted to a narrow range around $f=\epsilon=0.01$ because their sequential large pulses require a recovery time close to their characteristic time $\approx1/\epsilon=100$. In contrast, the P$n$ LCs can exist out to $f\approx 5\epsilon$ because their sequential pulses are subthreshold, thereby allowing the system to recover during the subthreshold sequence. When the subthreshold sequence duration extends to the recovery time $1/\epsilon$ the system is ready to produce the next superthreshold pulse resulting in the LC having $f\approx\epsilon$. In essence, both families of LCs are constrained to keep the period between large pulses close to its natural value $1/\epsilon$. The P2 LC provides the smooth transition from the period-$n$ LCs to the P$n$ LCs because it is a member of both families. 

The regime maps in Figs.\ \ref{fc-regime map} and \ref{fc-regime map zoom} suggest that by varying source amplitude at constant frequency, it is possible to interrogate exclusively either one of the two subharmonic LC families while crossing the island. This selectivity is seen in the $c$-continuation Poincar\'e plots in Fig.\ \ref{poincare-c diagrams} where only the period-$n$ LC family appears for $f=0.01$ and only P$n$ LCs appear for $f=0.04$.  

Throughout the island the locations of stable LCs and the regions of chaos between them are consistent with the LE plots and Poincar\'e bifurcation plots, thereby supporting the integrity of the numerical simulations used to create the map. Further confirmation of the numerically simulated regime map is provided by its similarity with the map generated from experimental measurements from the analog circuit in Fig.\ \ref{circuit}. Time series of voltages in Figs.\ \ref{screenshots} and \ref{per-n TS} measured from the circuit agree very well with their corresponding numerical predictions in Figs.\ \ref{f-contin}, \ref{time series P3}, \ref{time series P6 P8}, and \ref{continuation c=0.02 zm}. 

The two subharmonic LC families described above are the dominant LCs of the island. Other LCs exist as part of period-doubling cascades or as smaller periodic windows as seen within the chaos in the Poincar\'e and LE plots. These occurrences are not a subject of study here.   

Rajasekar and Lakshmanan did numerical investigations of chaos in the sinusoidally driven FHN oscillator using the source amplitude as a bifurcation parameter.\cite{rajasekar1988period} They found that the spacing convergence of the period-doubling bifurcation parameter values agreed with Feigenbaum's theory on the universal behavior of period-doubling systems.\cite{Feigenbaum1978,Feigenbaum1979} Their results are consistent with the period-doubling cascades to chaos seen in the right panel of Fig.\ \ref{poincare-c diagrams}. 

Sato and Doi applied periodic pulse stimulation to the fast variable's differential equation for the BVP oscillator.\cite{SATO1992243,doi1995global} They presented a regime map in the 2d parameter space of the pulse period and amplitude spanning the region that coincides with a map produced from measurements made on giant squid axon by N. Takahashi et.\ al.\cite{takahashi1990global} Agreement between Sato and Doi's numerically simulated map and Takahashi et.\ al.'s experimental map is very good. 

\begin{figure}[h]
    \centering
    \includegraphics[width=3 in]{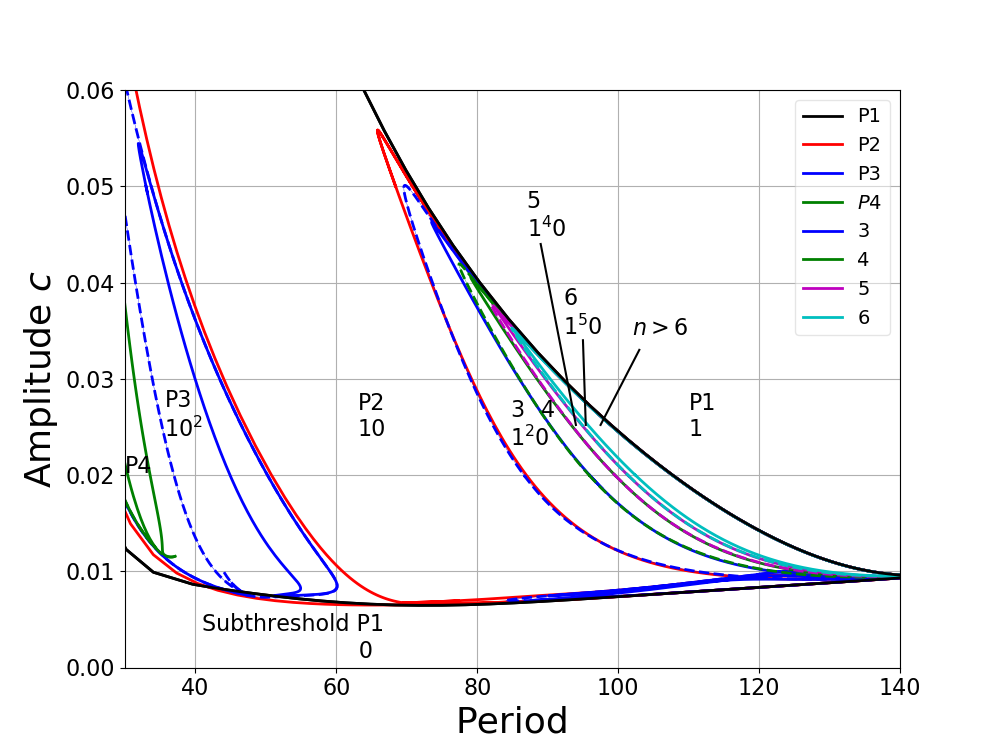}
    \caption{FHN simulation of regime map in the same range of parameter space used in Sato and Doi\cite{SATO1992243}. Examples of their LC notation are shown.}
    \label{per-c-map}
\end{figure}

Using the bifurcation methods applied here, we calculated the dynamics for Sato and Doi's BVP system and found that their map is a portion of an island of unstable P1 LC, consistent with the island in Fig.\ \ref{fc-regime map}. Their BVP system has the same progression as Fig.\ \ref{fc-regime map}, from superthreshold P1 to the period-$n$ LCs sequentially decreasing to a wide region of P2 LC, followed by the P$n$ LCs separated by chaotic regions, and finally ending in a reverse-doubling to subthreshold P1. Next, we compared Sato and Doi's map with our simulations of the driven FHN system by replotting Figs.\ \ref{fc-regime map} and \ref{fc-regime map zoom} using Sato and Doi's choice of axes. Figure \ref{per-c-map} shows the resulting map  
including examples of their notation. For example, $1^20$ means two large pulses followed by a small pulse, which is the same as the period-3 LC designation used here. The $10^2$ is one large pulse followed by two small pulses, which is the P3 LC. The strong resemblance of Fig.\ \ref{per-c-map} to the maps in Sato and Doi\cite{SATO1992243} and Takahashi et.\ al.\cite{takahashi1990global} is clear. The agreement is not surprising since the BVP and FHN systems are fundamentally the same, based on a cubic term of the fast variable with threshold positive feedback and on linear inhibition from a slow variable produced by linear kinetics. 

\begin{figure}[h]
    \centering
    \includegraphics[width=3 in]{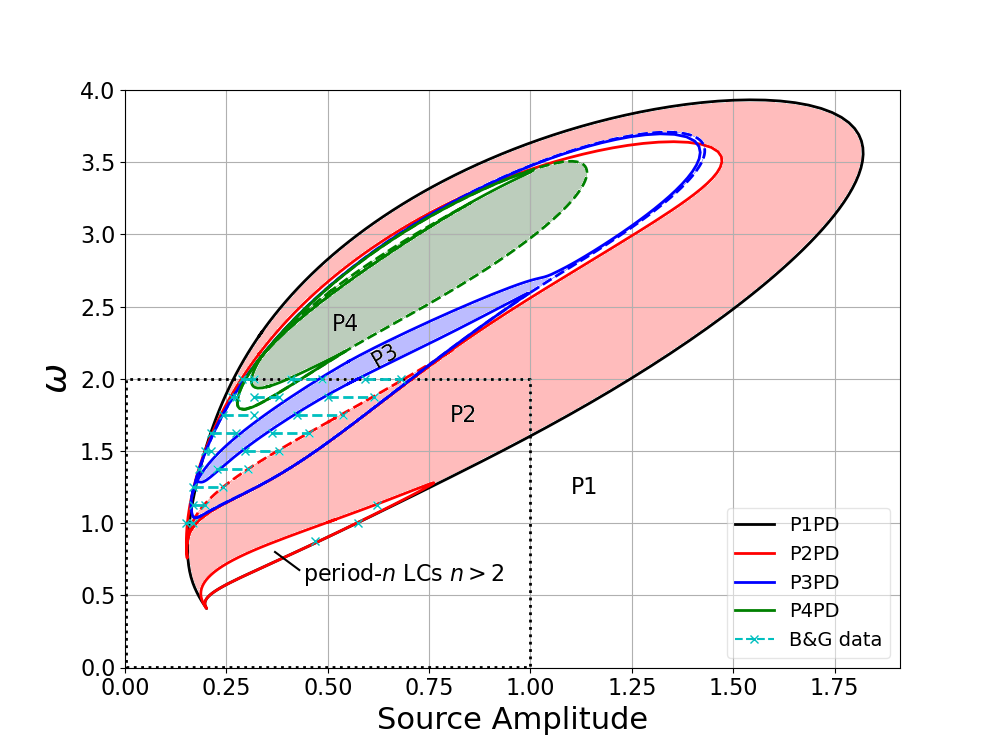}
    \caption{BVP simulation of regime map in source amplitude-frequency parameter space. Solid lines are PD, dashed are LP. Data from Barnes and Grimshaw is shown in teal. The choice of axes matches that used in Barnes and Grimshaw\cite{barnes1997numerical}.}
    \label{c-omega-map}
\end{figure}

Barnes and Grimshaw applied a sinusoidal source to the fast variable of the BVP oscillator.\cite{barnes1997numerical} They calculated the type of dynamics as stable limit cycle, quasi-periodic, or chaotic, over a grid of points in the 2d space of source amplitude and frequency. Their resulting regime maps identify regions where a stable LC exists and regions where no stable LCs exist. Using the bifurcation methods applied here on their BVP system, we produced the regime map in Fig.\ \ref{c-omega-map} using Barnes and Grimshaw's choice of axes  
for easy comparison to their figures. Included on the map are Barnes and Grimshaw's results which show that their unstable LC regions (teal) agree well with the chaotic regions between the P$n$ LCs. Note that their results exist only in the lower left quadrant of the parameter space indicated by the dotted lines and thus do not reveal the island nature of the region of unstable P1 LC. 

Coombes et.\ al.\ performed an analytical study of a piecewise linear version of the FHN which also show maps similar to Fig.\ \ref{per-c-map}.\cite{coombes2000period} Croisier and Dauby studied a FHN system similar to the one here where the stimulation was with periodic gaussian pulses.\cite{croisier2007continuation} Their period-continuation bifurcation plots clearly show the same two families of the dominant LCs seen in the frequency continuation plots in Figs.\ \ref{continuation c=0.02} and \ref{continuation c=0.02 zm}. 

Studies applying periodic stimulation to the slow variable have also been done.
Feingold et.\ al.\ did an investigation of the BVP system with pulse excitation applied to the slow variable.\cite{feingold1988phase} They produce a regime map in the 2d space of drive period and amplitude showing an unbounded region of unstable P1 LC. However, if a positive stimulation $c(1+\text{cos}(2\pi ft))$ is instead applied to the fast variable's differential equation, then the resulting map is an island of unstable P1 LC consistent with the island in maps presented here. The produced dynamics depend strongly on whether the forcing is applied in the fast or the slow variable's differential equation. 

The previously published studies described above used a BVP or FHN oscillator which was at rest when the periodic source amplitude was zero. If the system is self-firing when the source is zero ($0.1355<i<0.6168$ in Fig.\ \ref{i-contin}), then application of a periodic source causes dynamics very different from those presented here.\cite{CHOU1996,tsuji2004design}  These systems demonstrate quasi-periodic oscillations and can model action potential bursting. 

Recently, more complicated systems such as coupled oscillators, modified FHN oscillators, and control methods have been studied under the influence of periodic stimulation.\cite{zhang2023bifurcation,sakaguchi2023suppression,bosco2024influence} Understanding of these more complicated systems may benefit from the complex behavior displayed by the simpler FHN system presented here. 

\section{Conclusion}
For many decades the FHN system has been a popular simple model for the action potentials which occur across cell membranes of neural and muscle cells. An early area of investigation was the response of the FHN system to a periodic source. Period-adding LCs including period-doubling cascades to chaos were found. Bifurcation continuation plots and Poincar\'e plots showed the dependence of the dynamics on the source frequency and amplitude. Perhaps the most effective way to present the dynamics of the periodically driven FHN system is a 2d regime map in the parameter space of the source's frequency and amplitude. However, the previous studies could not benefit from today's computing capabilities and analysis techniques, which resulted in publication of limited portions of the regime map. Here a more comprehensive and detailed map is presented which reveals an island where LCs with the source's frequency are unstable. The island is dominated by two distinct families of subharmonic LCs and regions of chaos. The calculated map is shown to be consistent with the previously published limited maps. It also agrees well with a regime map constructed from measurements made from an analog circuit designed to calculate the FHN equations. The regime map presented here should contribute to a better understanding of the dynamics of the periodically driven FHN oscillator and help guide the design of future experiments using this popular excitable system. 

\bibliography{fhn_drvn-ms}

@PREAMBLE{
 "\providecommand{\noopsort}[1]{}" 
 # "\providecommand{\singleletter}[1]{#1}%" 
}

@article{FitzHugh1961,
title = {Impulses and Physiological States in Theoretical Models of Nerve Membrane},
journal = {Biophysical Journal},
volume = {1},
number = {6},
pages = {445-466},
year = {1961},
issn = {0006-3495},
doi = {https://doi.org/10.1016/S0006-3495(61)86902-6},
author = {FitzHugh, Richard},
}

@article{nagumo1962active,
  title={An active pulse transmission line simulating nerve axon},
  author={Nagumo, Jinichi and Arimoto, Suguru and Yoshizawa, Shuji},
  journal={Proceedings of the IRE},
  volume={50},
  number={10},
  pages={2061--2070},
  year={1962},
  publisher={IEEE}
}

@article{Feigenbaum1978,
  title = {Quantitative universality for a class of nonlinear transformations},
  author = {Feigenbaum, M.J.},
  journal = {J Stat Phys},
  volume = {19},
  pages = {25--52},
  numpages = {0},
  year = {1978},
  month = {Jul},
  doi = {10.1007/BF01020332},
  url = {https://doi.org/10.1007/BF01020332}
}

@article{Feigenbaum1979,
  title = {The universal metric properties of nonlinear transformations},
  author = {Feigenbaum, M.J.},
  journal = {J Stat Phys},
  volume = {21},
  pages = {669--706},
  numpages = {0},
  year = {1979},
  month = {Dec},
  doi = {10.1007/BF01107909},
  url = {https://doi.org/10.1007/BF01107909}
}

@article{cebrian2024six,
  title={Six decades of the FitzHugh--Nagumo model: A guide through its spatio-temporal dynamics and influence across disciplines},
  author={Cebri{\'a}n-Lacasa, Daniel and Parra-Rivas, Pedro and Ruiz-Reyn{\'e}s, Daniel and Gelens, Lendert},
  journal={Physics Reports},
  volume={1096},
  pages={1--39},
  year={2024},
  publisher={Elsevier}
}

@article{hodgkin1952quantitative,
  title={A quantitative description of membrane current and its application to conduction and excitation in nerve},
  author={Hodgkin, Alan L and Huxley, Andrew F},
  journal={The Journal of physiology},
  volume={117},
  number={4},
  pages={500},
  year={1952}
}

@article{van1927frequency,
  title={Frequency demultiplication},
  author={Van der Pol, Balth and Van Der Mark, Jan},
  journal={Nature},
  volume={120},
  number={3019},
  pages={363--364},
  year={1927},
  publisher={Nature Publishing Group UK London}
}

@article{rajasekar1988period,
  title={Period doubling route to chaos for a BVP oscillator with periodic external force},
  author={Rajasekar, S and Lakshmanan, M},
  journal={Journal of theoretical biology},
  volume={133},
  number={4},
  pages={473--477},
  year={1988},
  publisher={Elsevier}
}

@article{feingold1988phase,
  title={Phase locking, period doubling, and chaotic phenomena in externally driven excitable systems},
  author={Feingold, Mario and Gonzalez, Diego L and Piro, Oreste and Viturro, Hector},
  journal={Physical Review A},
  volume={37},
  number={10},
  pages={4060},
  year={1988},
  publisher={APS}
}

@article{Alexander1990,
author = {Alexander, James C. and Doedel, Eusebius J. and Othmer, Hans G.},
title = {On the Resonance Structure in a Forced Excitable System},
journal = {SIAM Journal on Applied Mathematics},
volume = {50},
number = {5},
pages = {1373-1418},
year = {1990},
doi = {10.1137/0150082},
URL = {https://doi.org/10.1137/0150082},
eprint = {https://doi.org/10.1137/0150082},
}

@article{takahashi1990global,
  title={Global bifurcation structure in periodically stimulated giant axons of squid},
  author={Takahashi, Nobuyuki and Hanyu, Yoshiro and Musha, Toshimitsu and Kubo, Ryogo and Matsumoto, Gen},
  journal={Physica D: Nonlinear Phenomena},
  volume={43},
  number={2-3},
  pages={318--334},
  year={1990},
  publisher={Elsevier}
}

@article{SATO1992243,
title = {Response characteristics of the BVP neuron model to periodic pulse inputs},
journal = {Mathematical Biosciences},
volume = {112},
number = {2},
pages = {243-259},
year = {1992},
issn = {0025-5564},
doi = {https://doi.org/10.1016/0025-5564(92)90026-S},
url = {https://www.sciencedirect.com/science/article/pii/002555649290026S},
author = {Shunsuke Sato and Shinji Doi}
}

@article{nomura1993bonhoeffer,
  title={A Bonhoeffer-van der Pol oscillator model of locked and non-locked behaviors of living pacemaker neurons},
  author={Nomura, Taishin and Sato, Shunsuke and Doi, Shinji and Segundo, Jose P and Stiber, Michael D},
  journal={Biological cybernetics},
  volume={69},
  number={5},
  pages={429--437},
  year={1993},
  publisher={Springer}
}

@article{Moss1994StochRes,
  title = {Stochastic resonance on a circle},
  author = {Wiesenfeld, Kurt and Pierson, David and Pantazelou, Eleni and Dames, Chris and Moss, Frank},
  journal = {Phys. Rev. Lett.},
  volume = {72},
  issue = {14},
  pages = {2125--2129},
  numpages = {0},
  year = {1994},
  month = {Apr},
  publisher = {American Physical Society},
  doi = {10.1103/PhysRevLett.72.2125},
  url = {https://link.aps.org/doi/10.1103/PhysRevLett.72.2125}
}

@article{doi1995global,
  title={The global bifurcation structure of the BVP neuronal model driven by periodic pulse trains},
  author={Doi, Shinji and Sato, Shunsuke},
  journal={Mathematical Biosciences},
  volume={125},
  number={2},
  pages={229--250},
  year={1995},
  publisher={Elsevier}
}

@article{BROWN1995359,
title = {Dynamic equilibria and oscillations of a periodically stimulated excitable system},
journal = {Chaos, Solitons \& Fractals},
volume = {5},
number = {3},
pages = {359-369},
year = {1995},
note = {Nonlinear Phenomena in Excitable Physiological Systems},
issn = {0960-0779},
doi = {https://doi.org/10.1016/0960-0779(93)E0028-A},
url = {https://www.sciencedirect.com/science/article/pii/0960077993E0028A},
author = {David Brown and Jonathan P.A. Foweraker and Robert W. Marrs},
}

@article{CHOU1996,
title = {Exotic dynamic behavior of the forced FitzHugh-Nagumo equations},
journal = {Computers \& Mathematics with Applications},
volume = {32},
number = {10},
pages = {109-124},
year = {1996},
issn = {0898-1221},
doi = {https://doi.org/10.1016/S0898-1221(96)00189-7},
url = {https://www.sciencedirect.com/science/article/pii/S0898122196001897},
author = {Mo-Hong Chou and Yu-Tuan Lin}
}

@article{Jurgen1997,
  title = {Coherence Resonance in a Noise-Driven Excitable System},
  author = {Pikovsky, Arkady S. and Kurths, J\"urgen},
  journal = {Phys. Rev. Lett.},
  volume = {78},
  issue = {5},
  pages = {775--778},
  numpages = {0},
  year = {1997},
  month = {Feb},
  publisher = {American Physical Society},
  doi = {10.1103/PhysRevLett.78.775},
  url = {https://link.aps.org/doi/10.1103/PhysRevLett.78.775}
}

@article{rajasekar1997control,
  title={Control of chaos by nonfeedback methods in a simple electronic circuit system and the FitzHugh-Nagumo equation},
  author={Rajasekar, S and Murali, K and Lakshmanan, M},
  journal={Chaos, Solitons \& Fractals},
  volume={8},
  number={9},
  pages={1545--1558},
  year={1997},
  publisher={Elsevier}
}

@article{barnes1997numerical,
  title={Numerical studies of the periodically forced Bonhoeffer van der Pol system},
  author={Barnes, Belinda and Grimshaw, Roger},
  journal={International journal of bifurcation and chaos},
  volume={7},
  number={12},
  pages={2653--2689},
  year={1997},
  publisher={World Scientific}
}

@article{coombes2000period,
  title={Period-adding bifurcations and chaos in a periodically stimulated excitable neural relaxation oscillator},
  author={Coombes, S and Osbaldestin, Andrew H},
  journal={Physical Review E},
  volume={62},
  number={3},
  pages={4057},
  year={2000},
  publisher={APS}
}

@article{tsuji2004design,
  title={A design method of bursting using two-parameter bifurcation diagrams in FitzHugh--Nagumo model},
  author={Tsuji, Shigeki and Ueta, Tetsushi and Kawakami, Hiroshi and Aihara, Kazuyuki},
  journal={International Journal of Bifurcation and Chaos},
  volume={14},
  number={07},
  pages={2241--2252},
  year={2004},
  publisher={World Scientific}
}

@article{kostova2004fitzhugh,
  title={FitzHugh--Nagumo revisited: Types of bifurcations, periodical forcing and stability regions by a Lyapunov functional},
  author={Kostova, Tanya and Ravindran, Renuka and Schonbek, Maria},
  journal={International journal of bifurcation and chaos},
  volume={14},
  number={03},
  pages={913--925},
  year={2004},
  publisher={World Scientific}
}

@article{volkov2005,
    author = {Volkov, E. I. and Ullner, E. and Kurths, J.},
    title = {Stochastic multiresonance in the coupled relaxation oscillators},
    journal = {Chaos: An Interdisciplinary Journal of Nonlinear Science},
    volume = {15},
    number = {2},
    pages = {023105},
    year = {2005},
    month = {04},
    issn = {1054-1500},
    doi = {10.1063/1.1899287},
    url = {https://doi.org/10.1063/1.1899287},
    eprint = {https://pubs.aip.org/aip/cha/article-pdf/doi/10.1063/1.1899287/14598155/023105\_1\_online.pdf},
}

@article{croisier2007continuation,
  title={Continuation and bifurcation analysis of a periodically forced excitable system},
  author={Croisier, Huguette and Dauby, PC},
  journal={Journal of theoretical biology},
  volume={246},
  number={3},
  pages={430--448},
  year={2007},
  publisher={Elsevier}
}

@article{croisier2009bifurcation,
  title={Bifurcation analysis of a periodically forced relaxation oscillator: Differential model versus phase-resetting map},
  author={Croisier, H and Guevara, MR and Dauby, PC},
  journal={Physical Review E—Statistical, Nonlinear, and Soft Matter Physics},
  volume={79},
  number={1},
  pages={016209},
  year={2009},
  publisher={APS}
}

@article{shimizu2012complex,
  title={Complex mixed-mode oscillations in a Bonhoeffer--van der Pol oscillator under weak periodic perturbation},
  author={Shimizu, Kuniyasu and Saito, Yuto and Sekikawa, Munehisa and Inaba, Naohiko},
  journal={Physica D: Nonlinear Phenomena},
  volume={241},
  number={18},
  pages={1518--1526},
  year={2012},
  publisher={Elsevier}
}

@article{takahashi2018mixed,
  title={Mixed-mode oscillation-incrementing bifurcations and a devil’s staircase from a nonautonomous, constrained Bonhoeffer--van der Pol oscillator},
  author={Takahashi, Hiroaki and Kousaka, Takuji and Asahara, Hiroyuki and Stankevich, Nataliya and Inaba, Naohiko},
  journal={Progress of Theoretical and Experimental Physics},
  volume={2018},
  number={10},
  pages={103A02},
  year={2018},
  publisher={Oxford University Press}
}

@article{inaba2020nested,
  title={Nested mixed-mode oscillations, part II: Experimental and numerical study of a classical Bonhoeffer--van der Pol oscillator},
  author={Inaba, Naohiko and Tsubone, Tadashi},
  journal={Physica D: Nonlinear Phenomena},
  volume={406},
  pages={132493},
  year={2020},
  publisher={Elsevier}
}

@article{zhang2023bifurcation,
  title={Bifurcation analysis of a modified FitzHugh-Nagumo neuron with electric field},
  author={Zhang, Xu and Min, Fuhong and Dou, Yiping and Xu, Yeyin},
  journal={Chaos, Solitons \& Fractals},
  volume={170},
  pages={113415},
  year={2023},
  publisher={Elsevier}
}

@article{sakaguchi2023suppression,
  title={Suppression and frequency control of repetitive spiking in the FitzHugh-Nagumo model},
  author={Sakaguchi, Hidetsugu and Yamasaki, Keito},
  journal={Physical Review E},
  volume={108},
  number={1},
  pages={014207},
  year={2023},
  publisher={APS}
}

@article{bosco2024influence,
  title={Influence of sinusoidal forcing on the master FitzHugh--Nagumo neuron model and global dynamics of a unidirectionally coupled two-neuron system},
  author={Bosco, N{\'\i}vea D and Rech, Paulo C and Beims, Marcus W and Manchein, Cesar},
  journal={Chaos: An Interdisciplinary Journal of Nonlinear Science},
  volume={34},
  number={9},
  year={2024},
  publisher={AIP Publishing}
}

@article{inaba2024nested,
  title={Nested mixed-mode oscillations in the forced van der Pol oscillator},
  author={Inaba, Naohiko and Okazaki, Hideaki and Ito, Hidetaka},
  journal={Communications in Nonlinear Science and Numerical Simulation},
  volume={133},
  pages={107932},
  year={2024},
  publisher={Elsevier}
}

@article{wolf1985,
title = {Determining Lyapunov exponents from a time series},
journal = {Physica D: Nonlinear Phenomena},
volume = {16},
number = {3},
pages = {285-317},
year = {1985},
issn = {0167-2789},
doi = {https://doi.org/10.1016/0167-2789(85)90011-9},
url = {https://www.sciencedirect.com/science/article/pii/0167278985900119},
author = {Alan Wolf and Jack B. Swift and Harry L. Swinney and John A. Vastano}
}

@book{ermentrout,
  title={Simulating, Analyzing, and Animating Dynamical Systems: A Guide to XPPAUT for Researchers and Students},
  author={Ermentrout, B.},
  isbn={9780898715064},
  series={Software, Environments and Tools (Book 14)},
  year={2002},
  publisher={SIAM}
}

@Misc{auto-07p,
OPTkey = {•},
author = {Doedel, Eusebius J. and Fairgrieve, Thomas F. and Sandstede, Björn and Champneys, Alan R.
and Kuznetsov, Yuri A. and Wang, Xianjun},
title = {AUTO-07P: Continuation and bifurcation software for ordinary differential equations},
howpublished = {Concordia University},
OPTmonth = {•},
year = {2012},
OPTnote = {•},
OPTannote = {•}
}
\end{document}